# Job insecurity and equilibrium determinacy in a rational expectations, New Keynesian model with asymmetric information. A theoretical analysis

Luca Vota[1*], Luisa Errichiello[2]

## Abstract

Despite the importance of this variable in the macroeconomic context, current research on job insecurity remains mainly confined to its non-systemic dimension. The research aim of this paper is to identify the short-run and long-run macroeconomic determinants of job insecurity in the presence of asymmetric information between public and private agents, informative shocks, and different degrees of institutional communication transparency. To accomplish this goal, a small-scale, rational expectations, New Keynesian model is proposed in which limitedly informed households and firms receive a potentially noisy informative signal about the unobservables from fully informed government and central bank. It is found that, notwithstanding the fulfillment of the Taylor principle, if public agents transfer all the available information to the private agents without communication ambiguities, the model admits a unique, stable equilibrium path along which the "Paradox of Transparency" can emerge. Otherwise, the model's dynamics become unpredictable in terms of equilibrium existence and multiplicity, and job insecurity plays a potentially fundamental role in equilibrium determinacy. Appropriate policy recommendations are discussed.

## 1. Introduction

Precarious work is a particularly widespread phenomenon globally, both in advanced economies and in emerging and developing countries. According to a survey conducted by the Ipsos institute on 12,000 adult workers residing in 27 countries and presented at the World Economic Forum in 2020 (Ipsos 2020), 54% of the sample declared themselves concerned about the possibility of losing their job in the 12 months following the interview (of these, 17% defined themselves as "very concerned", while 37% said they were "fairly concerned"). The highest values were recorded in Russia (75%), Spain (73%), and Malaysia (71%), while the lowest ones were reported in Germany (26%), Sweden (30%), the Netherlands, and the United States (36%). Although they indicate strong variability between countries, these data demonstrate that job insecurity, defined as the subjectively perceived probability of experiencing an interruption in one's career (Shoss, 2017), is a really pervasive issue even in countries with a strong industrial vocation that offer their workers generous welfare measures (as in the case of Germany and Sweden).

Starting from the 1990s, precisely because of the structural changes that have affected economies globally and that have made traditional relationships between workers and businesses less structured and more flexible (Davis, 2013), social scientists have shown a growing interest in job insecurity. Factors such as globalization and automation have made a decisive contribution to reducing the costs and times of production activities but, at the same time, have increased workers' perception of insecurity (Mughan et al., 2003; Scheve & Slaughter, 2006; Couple, 2019; Nam, 2019; Raeder et al., 2019; Cao & Song, 2025). Contrary to what one might imagine, technological progress does not constitute a threat solely to the career continuity of low-skilled workers, but also to that of individuals

---
[1*] Institute for Studies on the Mediterranean of the National Research Council (CNR-ISMed), Naples, Italy.
luca.vota@cnr.it
[2] Institute for Studies on the Mediterranean of the National Research Council (CNR-ISMed), Naples, Italy.
luisa.errichiello@cnr.it

with high professional qualifications (Colvin, 2015). More recent evidence produced from qualitative and quantitative studies has highlighted that the introduction of Artificial Intelligence (AI) has significantly influenced perceived job insecurity, without relevant differences between employees in managerial roles and those in non-managerial roles (Koo et al., 2021). Others, however, have discovered that the negative impact of AI on job insecurity is mediated by vocational learning ability, i.e., the ability to autonomously acquire and apply new skills to one's own work environment (Liu & Zhan, 2020). In fact, the need to interact with AI to perform their duties encourages workers to constantly seek to improve their skills, with important implications for their insecurity, creativity, well-being, and psychological health (Wu et al., 2024). The fear that AI will rapidly make their skills obsolete, leading to the replacement of human operators with virtual ones, is another important source of stress for workers (Sharif et al., 2025).

Existing research to date has favored a micro and meso approach focused on individual, job-related and organizational antecedents of job insecurity, paying little attention to systemic, macro-economic ones. In particular, they have used survey data relating to variables such as contract type, responsibilities associated with one's position, hourly wage, health conditions, number of children, age, years of experience, and so on.

Such analyses have often focused on specific sectors and regions to account for the heterogeneity present in different economic sectors and in the economic conditions of various countries, proving useful in evaluating the predictors of job insecurity in different contexts (Lee et al., 2018; Chirumbolo et al., 2020; Martínez et al., 2020; De Cuyper et al., 2021; Ibanescu et al., 2023; Muñoz Medina et al., 2023; Darvishmotevali, 2025).

However, a fundamental limitation of the individual-based approach is its inability to account for macroeconomic variables and (consequently) the influence that the alternation of the various phases of the business cycle exerts on job insecurity. Indeed, while purely individual variables undoubtedly have a significant impact on job insecurity, it is equally plausible that a non-negligible component of insecurity is predicted by macroeconomic variables such as actual GDP, output gap, inflation, unemployment rate, investment, household consumption and saving, and so on. As demonstrated by the literature on Dynamic Stochastic General Equilibrium (DSGE) models, these variables move along one or multiple equilibrium paths in response to exogenous shocks that are more or less persistent and potentially capable of triggering phases of expansion and recession (Dave & Sorge, 2025). Governments and central banks react to cyclical fluctuations by using the economic policy tools at their disposal (taxation, public spending, social transfers, interest rates, money supply) to stabilize price and national income growth, preserve employment, and contain the harmful effects of adverse shocks. This complex macroeconomic dynamics inevitably reflects on the expectations of households and firms about the fundamentals and future perspectives of being unemployed.

The lack of studies focused on the macroeconomic dimension does not allow for discerning the short-run and long-run effects of exogenous shocks on job insecurity, nor for defining public policies to combat job insecurity that adequately account for the effects of the business cycle. In fact, to effectively design and implement their policies for containing job insecurity, policymakers need analytical tools that consider the gradual processes of divergence and convergence of the actual unemployment rate with respect to its steady-state equilibrium (which, in turn, are strictly dependent on cyclical fluctuations).

This manuscript attempts to fill this gap by proposing a rational expectations New Keynesian model populated by public agents (central bank and government) who are fully informed and private agents (firms and households) who directly observe only a part of the state variables and who receive a potentially incomplete and noisy informative signal regarding unobservables and informative shocks (which is used by them to formulate their long-run expectations about the fundamentals). The expression "potentially incomplete and noisy" indicates that the information transmitted by public agents to private ones can be partial (i.e., not include all values of the unobservables) and flawed by ambiguity.

The main innovation of the work consists in demonstrating that adopting a fully transparent communication strategy represents a sufficient condition to guarantee equilibrium determinacy and uniqueness, regardless of whether the Taylor Principle is satisfied (i.e., the central bank responds aggressively to an increase in prices by raising its policy rate more than proportionally to the inflation rate). However, along this equilibrium path, the "Paradox of Transparency'" can emerge, meaning that the transmission of otherwise unobservable information by public agents to private ones can increase job insecurity and lead to a loss in terms of welfare (Morris & Shin, 2005).

The rest of the paper is organized as follows: Section 2 proposes a literature review on job insecurity, with a clear indication of the gap filled by this research work; Section 3 presents the rational expectations New Keynesian model with asymmetric information; Section 4 lists the common knowledge solution of the model; Section 5 illustrates the model's asymmetric information characterization; Section 6 discusses the obtained theoretical results; Section 7 proposes an extension of the theoretical analysis for the aggregate economy to the specific but relevant context of mature workers; finally, Section 8 reports the concluding remarks of the study. A detailed bibliography closes the paper.

## 2. Job insecurity research: moving forwards a systemic approach

At its core, job insecurity is the perceived threat that one's job will not continue, a forward-looking construct widely examined across sociology, economics, and organizational psychology (Heaney, Israel, & House, 1994; Shoss, 2017). By distinguishing these perceptions from realized labor-market events, the literature lays the groundwork for more fine-grained distinctions.

First, objective vs. subjective job (in)security differentiates realized or externally observed risk (e.g., contract type, displacement) from personal expectations and worries about job continuity (De Witte & Näswall, 2003; Erlinghagen, 2008). Second, quantitative vs. qualitative insecurity distinguishes perceived threats to the job itself (quantitative) from threats to valued features of the job (qualitative) - such as career prospects, learning opportunities, pay growth - (Hellgren et al., 1999; Urbanaviciute et al., 2021). Third, cognitive vs. affective insecurity separates likelihood judgments ("How likely am I to lose my job?") from emotional reactions ("How worried am I?") (Borg & Elizur, 1992; Huang et al., 2012).

Measurement mirrors these distinctions. Objective indicators include contract status (temporary/permanent) and unemployment spells and displacement, and institutional features like employment protection. These are prevalent in economic analyses linking structural conditions to security outcomes whereas macro-comparative work often aggregates objective proxies at country level (Burgard et al., 2009; De Witte & Näswall, 2003; Green & Leeves, 2013). Subjective insecurity, widely used in organizational research and sociology, overwhelmingly use perceptual instruments and job (in)security is usually captured via multi-item survey scales, typically use Likert-type items (e.g., "I might lose my job in the next year") (e.g., Ashford et al., 1989; Hellgren, Sverke, & Isaksson, 1999; Vander Elst et al., 2014) and many have undergone reliability and invariance testing across languages and countries (Lee at al., 2008; Görgens-Ekermans et al., 2024).

Within organizational psychology and sociology traditions, a large body of work documents antecedents and consequences of subjective job insecurity. Meta-analytic evidence shows that job insecurity is linked to poorer mental and physical health, reduced job satisfaction and commitment, higher turnover intentions, and, in many cases, diminished performance (Sverke et al., 2002; Sverke et al., 2019; Hur, 2022).

Antecedents operate at multiple levels. Across various literature reviews (Keim et al., 2014; Shoss, 2017; Lee, Huang & Ashford, 2018; Jian et al., 2021), they cluster into: institutional/macro factors (e.g., labor-market institutions, unemployment); organizational factors (e.g., restructuring/downsizing, union presence); job/position factors (e.g., temporary or contingent contracts, part-time status, lower-skill/blue-collar roles); individual demographics (e.g., tenure, minority status, education); individual experiential factors (e.g., perceived employability, training, prior unemployment); personality/core self-evaluations (e.g., locus of control, affectivity, self-esteem); and interpersonal/social factors (e.g., bullying, leadership, coworker/supervisor support). Focusing on Jiang et al. (2021), their multilevel meta-analysis adopts a Job Demands–Resources lens, recasting determinants as resources (personal and organizational) that reduce job insecurity and demands (personal and organizational) that increase it - showing that resources exert the stronger overall effects, underscoring the protective impact of employability, fair procedures, participation, and supportive leadership.

Cross-disciplinary studies also highlight that subjective perceptions are not epiphenomenal: they only partly overlap with objective conditions and can exert independent effects on well-being and behavior (De Witte & Näswall, 2003; Burgard et al., 2009). For example, temporary work does not uniformly depress attitudes, whereas subjective insecurity robustly predicts lower satisfaction and commitment across institutional settings (De Witte & Näswall, 2003). At the same time, research has begun to differentiate cognitive and affective facets, with the latter (worry) sometimes mediating or amplifying the effects of the former (likelihood judgments) on strain and performance (Jiang & Lavaysse, 2018; Huang et al., 2012). Relatedly, literature on employability suggests that perceived external options can buffer insecurity's effects - though not uniformly across contexts or populations (Fugate et al., 2004; de Cuyper et al., 2014; Yeves et al., 2019).

A second, largely sociological and economic streams scales up the analysis to the country level, examining whether macro-institutions and context shape average insecurity levels and their distribution. Drawing on extant research, Errichiello & Falavigna (2024) propose a framework grouping environmental factors shaping job insecurity into two broad groups: institutional categories and socio-economic structural factors. Institutions cover the following areas: (i) labour market; ii) social protection; iii) worker's; iv) national culture; and, v) governance. Studies leverage correlational, multilevel models along with secondary data and cross-national surveys to relate insecurity to: active labour market policies (ALMPs) (e.g., Chung & Van Oorschot, 2011; Inanc & Kalleberg, 2022; Lübke & Erlinghagen, 2014); employment protection legislation (EPL) (e.g., Balz, 2017; Inanc & Kalleberg, 2022); social protection measures (e.g., Erlinghagen, 2008; Inanc & Kalleberg, 2022), worker's power (union density and collective bargaining) (Esser & Olsen, 2012; Inanc & Kalleberg, 2022), national cultural values (Erlinghagen, 2008; Moy et al., 2023; Errichiello & Falavigna, 2024), and quality of governance (e.g., Dixon et al., 2013; Errichiello & Falavigna, 2024) As for structural factors related to labour maket and socio-economic conditions, empirical research correlate JI to both current and long-term high unemployment (e.g., Chung & van Oorschot, 2011), GDP growth rate (e.g., Lübke & Erlinghagen, 2014) and country-level income inequality (Errichiello & Falavigna, 2025). This work typically compares countries or time periods around crises (e.g., 2008), reporting that macroeconomic slack, institutional protections, and cultural-institutional contexts correlate with perceived insecurity - yet with mixed and sometimes modest effect sizes across outcomes and subpopulations.

Despite their contributions, existing studies exhibit two limitations germane to the present paper's aims. First, most are correlational and rely on static or quasi-static designs (cross-sectional snapshots or pooled panels). This makes it difficult to model transmission mechanisms - for example, how shocks to aggregate demand, policy interventions (ALMPs, EPL reforms), or changes in governance quality propagate into workers' expectations about job continuity over time. Second, even when macro variables are included, these models rarely have explicit predictive structure: they do not incorporate expectation formation, feedback loops between unemployment/output and perceptions, or the role of policy rules, all central to New-Keynesian (NK) frameworks. Consequently, they are limited in forecasting subjective job insecurity under counterfactual monetary-fiscal or ALMP/EPL scenarios, or in evaluating the dynamic welfare and distributional consequences of policy regimes. These gaps are consequential because subjective quantitative insecurity is precisely where macro shocks meet micro behavior. Recent reviews argue that insecurity is shaped by both demands (e.g., reorganization, role overload, macro slack) and resources (e.g., communication, employability, institutional buffers), with resources often exerting stronger protective effects (Jiang et al., 2021). Embedding this insight in a macro model requires a structural mapping from aggregate states (output gap, unemployment, policy rates, inflation, fiscal stance, etc.) to agents' expectations of job loss, allowing those expectations to influence consumption, labor supply, and wage setting - canonical NK channels. Doing so would extend macro-comparative work that links insecurity to structural socio-economic factors and institutions (e.g., Erlinghagen, 2008; Chung & van Oorschot, 2011; Lübke & Erlinghagen, 2014) by providing a dynamic, forward-looking, and policy-counterfactual apparatus that can forecast trajectories of subjective risk, not just correlate them with realized outcomes.

## 3. An original New Keynesian model

Consider an economy populated by private agents (households and firms) who optimize their respective expected profit and expected utility functions, a central bank that has a dual mandate (to stabilize the actual inflation rate and minimize the output gap), and a fiscal policy authority (government) that collects resources through taxation and re-employs them through government spending.

It is assumed that the central bank and the government are fully informed about macroeconomic variables (actual output, actual inflation, potential output, consumption, saving, investment, public spending, taxes, effective interest rate, natural interest rate, and unemployment rate) and exogenous shocks (shock to potential output, cost-push shock to inflation, informative shock, shock to consumer preferences, shock to natural unemployment, fiscal shocks, idiosyncratic shock) affecting the economy, while private agents can directly observe the current values of the actual interest rate ($r_t$), current inflation ($\pi_t$), and all past values of the model's endogenous variables (actual output, consumption, investment, output gap, central bank nominal interest rate, saving, effective interest rate, inflation rate, unemployment rate).

It is also assumed that in this economy there is an informative signal composed of two parts. The first is an observable component consisting of information about unobservables that are directly disclosed by the central bank and the government ($\chi_t$) and unexpected informative shocks, which is the part of information that public agents generally do not disclose during their ordinary communication activities ($\lambda_t$). The second, instead, is the noisy component of the informative signal (hereinafter indicated as $\Xi_{t|t}$) which is generated by the ambiguity of the public agents' communication strategy and cannot be directly observed by private agents (Gabaix, 2020).

In light of the above, the overall informative signal $\Psi_{t|t}$ can be defined as follows (Morris & Shin, 2002; Jarociński & Karadi, 2020):

$$\Psi_{t|t} = \chi_t + \Xi_{t|t} \tag{1}$$

where $\Xi_{t|t}$ is a white noise process, whereas $\chi_t$ evolves according to a stationary AR(1) process (Blanchard et al., 2013; Blinder et al., 2024):

$$\chi_t = \rho_\chi \chi_{t-1} + \lambda_t \tag{2}$$

namely, $|\rho_\chi| < 1$. The overall informative signal $\Psi_{t|t}$ is naturally not directly observable from the private sector's side, because it is given by the sum between an observable component ($\chi_t$) and an unobservable and unpredictable one ($\Xi_{t|t}$).

The full information set is defined as as $\mathcal{F}^t = \{\mathcal{F}_t, \mathcal{F}_{t-1}, ... \}$, namely, respect to the entire history of the observables contained in $\mathcal{F}_t = \{y_{t|t}, \hat{y}_{t|t}, c_{t|t}, S_{t|t}, I_{t|t}, i_{t|t}, \pi_t, \bar{y}_{t|t}, u_{t|t}, g_{t|t}, tax_{t|t}, r_t, \bar{r}_{t|t}, \eta_{t|t}, v_{t|t}, \xi_{t|t}, \chi_t, \lambda_t, \omega_{t|t}, \epsilon_{t|t}, \varsigma_{t|t}, \mathfrak{T}_{t|t}, \varepsilon_{t|t}, \mathfrak{Q}_{t|t}, \mathfrak{y}_{t|t}, \Xi_{t|t}\}$, where the subscript $t|t$ indicates the variables that can be directly observed only by the central bank and the government, while the subscript $t$ labels the variables that can be directly observed by all the agents.

Let $\mathcal{H}^t = \{\mathcal{H}_t, \mathcal{H}_{t-1}, \mathcal{H}_{t-2}, \dots \mathcal{H}_{t-i}, \dots, \mathcal{H}_{t-I}\}$ be the private agents' information set, where $\mathcal{H}_{t-i}$ with $i = 0, \dots, I$ is a vector whose unique non-null elements are given by the actual interest rate $r_{t-i}$, inflation rate ($\pi_{t-i}$), informative shock ($\lambda_{t-i}$) and informative signal transmitted to households and firms by the public agents ($\chi_{t-i}$).

The hypothesis about the information distribution above is very common in the DSGE modelling (Lubik et al., 2023), because it consists of making the weak and realistic assumption that private agents need to be able to observe at least the prices in the economy (including the interest rate, of course) to make their allocative decisions.

Assume that the economy's structure is described by a small-scale New Keynesian model in its linearized form.

Following Galí et al. (2004), households do not set their current consumption level $c_{t|t}$ on the basis of the standard Euler equation but their expectations about long-term income conditional on their information set (Christiano et al., 2008; Blanchard et al., 2013), government spending (Ravn et al., 2006; Galí et al., 2007), and taxes (Blanchard & Perotti, 2002; Romer & Romer, 2010):

$$c_{t|t} = c_0 + c_1 \lim_{h \to +\infty} E_t\left[y_{t+h|t} | \mathcal{H}_t\right] + c_3 g_{t|t} - c_4 tax_{t|t} + \eta_{t|t} \tag{3}$$

In equation (3), that part of consumption that does not depend on expected income and is not part of the autonomous component ($c_0$), namely the error term $\eta_{t|t}$, is a linear combination of white noise demand shocks $\xi_{t|t}$ (which express changes in preferences or spending behaviors), the information transmitted by public agents to private ones $\chi_t$, and an idiosyncratic white noise shock $\upsilon_{t|t}$ which represents shocks affecting agents at the individual level (for example, an injury that may have consequences on job continuity, with negative consequences on the current consumption level):

$$\eta_{t|t} = \varphi_1 \xi_{t|t} + \varphi_2 \chi_t + \varphi_3 \upsilon_{t|t} \tag{4}$$

The modeling of shocks expressed by equation (4) is coherent with Smets & Wouters (2007) and An & Schorfheide (2007).

Since $\chi_t$ has an AR(1) representation, assumption (4) introduces persistence effects into the mechanism of information transmission from fully informed agents to limitedly informed ones (i.e., households and firms gradually incorporate into their information set the data communicated to them by the central bank and the government). This implies the existence of a time lag between the moment when news is disseminated (time $t$) and the moment when it is actually incorporated into households' consumption decisions (time $t + 1$).

Consistently with equation (3), current household saving also depends on their expectations regarding long-term income, the return on saving (i.e., the interest rate), and (once again) on the shock to preferences and news shock:

$$S_{t|t} = s_0 + s_1 \lim_{h \to +\infty} E_t\left[y_{t+h|t} | \mathcal{H}_t\right] + s_2 r_t - s_3 g_{t|t} - s_4 tax_{t|t} + \eta_{t|t} \tag{5}$$

This expression is consistent with the long-standing Keynes-Friedman-Modigliani and the saver-spender traditions, according to which saving is a linear function of disposable income, taxes, and the interest rate (Keynes, 1936; Friedman, 1957; Ando & Modigliani, 1963; Mankiw, 2000). An example of New Keynesian models incorporating a similar expression of saving is Galí et al. (2004). The contribution of Campbell & Mankiw (1989) provides further theoretical justification for this specification of saving in the context of New Keynesian modelling.

Firms' investment depends on expectations regarding long-term national income growth, the cost of the investment itself (i.e., the actual interest rate $r_t$), and the current information flow:

$$I_{t|t} = \gamma_1 \lim_{h \to +\infty} E_t[y_{t+h|t} - y_{t|t}|\mathcal{H}_t] - \gamma_2 r_t - \gamma_3 g_{t|t} - \gamma_4 tax_{t|t} + \gamma_5 \chi_t \qquad (6)$$

In equation (6), the coefficient $\gamma_3$ accounts for the potential crowding-out effect of government spending. This expression of investment is supported by the user cost theory (Jorgenson, 1963), reduced-form, linear Tobin's q (Tobin, 1969), and the fiscal literature (Hall & Jorgenson, 1967). More recently a similar specification has been used in a New Keynesian model by Galí et al. (2004) and Christiano et al. (2005).

According to equation (6), investment increases only if firms expect long-term national income growth ($y_{t+h|t} > y_{t|t}$). If, on the other hand, they expect a contraction in long-term national income ($y_{t+h|t} < y_{t|t}$), firms will disinvest.

The aggregate behavior of the private sector (households and firms) is expressed by the dynamic IS curve:

$$\hat{y}_{t|t} = E_t[\hat{y}_{t+1|t}|\mathcal{H}_t] - \frac{1}{\sigma}(i_{t|t} - E_t[\pi_{t+1}|\mathcal{H}_t] - \bar{r}_{t|t}) \qquad (7)$$

where $\hat{y}_{t|t}$ is the output gap, $\pi_t$ is the inflation rate, and $\bar{r}_{t|t}$ is the natural interest rate following the standard Wicksell equation:

$$\bar{r}_{t|t} = \sigma(E_t[\bar{y}_{t+1|t}|\mathcal{H}_t] - \bar{y}_{t|t}) \qquad (8)$$

where $\bar{y}_{t|t}$ is potential output.

The relationship between national income and unemployment rate is dictated by the Okun's Law as in Galí et al. (2012) and Ball et al. (2017):

$$u_{t|t} - \bar{u}_{t|t} = -\theta(y_{t|t} - \bar{y}_{t|t}) \qquad (9)$$

where the natural unemployment rate follows a stationary AR(1) process ($|\rho_u| < 1$) as in Basistha & Nelson (2007):

$$\bar{u}_{t|t} = \rho_u \bar{u}_{t-1|t} + \mathfrak{T}_{t|t}$$

Assume that households and firms, being endowed with rational expectations, know the structure of the economy, and in particular know that national income follows a random walk with time-varying drift $\mu_{t|t}$ (Fernald, 2014; Del Negro et al., 2015):

$$y_{t|t} = \mu_{t|t} + y_{t-1} + \varepsilon_{t|t} \qquad (10)$$

where, of course, $\mu_{t|t}$ is nothing but households' optimal estimate of potential GDP which, by assumption, follows a stationary AR(1) process (Sments & Wouters, 2007):

$$\mu_{t|t} = \rho_{\bar{y}} \mu_{t-1|t} + \omega_{t|t} \qquad (11)$$

where $|\rho_{\bar{y}}| < 1$, is an exogenous potential output shock and $\mu_{t-1} \equiv \bar{y}_{t-1}$, since potential GDP at time $t-1$ can be observed by all agents. Since private agents cannot observe $\omega_{t|t}$, they commit a white noise forecasting error which is indicated in equation (10) by $\varepsilon_{t|t}$.

The dynamics of inflation are described by the New Keynesian Phillips curve:

$$\pi_t = \beta E_t[\pi_{t+1}|\mathcal{H}_t] + k\hat{y}_{t|t} + \epsilon_{t|t} \tag{12}$$

where $\epsilon_{t|t}$ is a term evolving according to a stationary AR(1) process (Christiano et al., 2005):

$$\epsilon_{t|t} = \rho_\epsilon \epsilon_{t-1|t} + \varsigma_{t|t} \tag{13}$$

where $|\rho_\epsilon| < 1$ and $\varsigma_{t|t}$ è an exogenous cost-push shock to inflation.

The central bank has its nominal interest rate as its sole policy instrument, which is set based on a standard Taylor rule:

$$i_{t|t} = \alpha_\pi \pi_t + \alpha_y \hat{y}_{t|t} \tag{14}$$

Assume that both tax revenue and government spending follow, by assumption, a stationary AR(1) process (Leeper et al., 2010A), i.e.:

$$g_{t|t} = \rho_g g_{t-1|t} + \mathfrak{v}_{t|t} \tag{15}$$

And (Leeper et al., 2010B):

$$tax_{t|t} = \rho_{tax} tax_{t-1|t} + \mathfrak{L}_{t|t} \tag{16}$$

where $\mathfrak{L}_{t|t}$ and $\mathfrak{v}_{t|t}$ are exogenous, white noise fiscal shocks with $\left|\rho_g\right| < 1$, and $\left|\rho_{tax}\right| < 1$.

The macroeconomic equilibrium of this model (assuming at least one exists) is determined under the condition that the financial market is in equilibrium (i.e., that aggregate saving equals investment):

$$S_{t|t} = I_{t|t} \tag{17}$$

The macroeconomic resource constraint is another fundamental equilibrium condition of the model:

$$y_{t|t} = I_{t|t} + c_{t|t} + g_{t|t} \tag{18}$$

It is also assumed that, in equilibrium, the government's fiscal decisions are bound by the usual budget constraint:

$$g_{t|t} = tax_{t|t} \tag{19}$$

Although it may be a simplification compared to their real Data Generating Process (DGP), the assumption that macroeconomic shocks follow an AR(1) process is very common in the literature (Smets & Wouters, 2003; Blanchard & Galì, 2007; de Jesus et al., 2020), as it effectively balances the need to account for the persistence mechanisms of exogenous shocks with the need to avoid overly complex specifications that could make the entire model mathematically intractable. Furthermore, specific empirical analyses support the validity of this assumption for keynote estimated DSGEs (Peersman & Straub, 2006).

The solution of the model formed by equations (1)-(19) requires first identifying the possible full information equilibrium (i.e., that which would be obtained in the hypothetical situation where all agents have a complete information set). This solution is also referred to as the "common knowledge" one, as the full information situation is realistically achieved thanks to the public agents' disclosure process.

The common knowledge equilibrium is identified by solving the system formed by equations (1)-(19) where $j_{t|t} = j_t$ and $j_{t|t}$ lables the generic model's variable.

## 4. The common knowledge solution

The common knowledge solution of this model is determined by assuming that the central bank and the government disclose all data not included in the original information set of private agents (households and firms) without ambiguities or errors in their communication activities. Therefore, $\Xi_{t|t} = \Xi_t = 0$ and $\Psi_{t|t} = \Psi_t = \chi_t$, which means that the private agents' information set coincides with that of their public counterparts.

The equations of the reduced form of the common knowledge model are reported below. The generic coefficient $z_i^j$ that appears in each expression is a nonlinear combination of the parameters of the model's structural form. The values of all these coefficients are reported in the tables in the Appendix.

*Proposition 1*

Along the common knowledge equilibrium path, the actual interest rate $r_{t|t}^* = r_t^*$ is given by:

$$r_t^* = z_0^r + z_1^r \bar{y}_{t-2} + z_2^r \omega_{t-1} + z_3^r g_{t-1} + z_4^r \mathfrak{y}_t + z_5^r tax_{t-1} + z_6^r \mathfrak{Q}_t + z_7^r \chi_{t-1} + z_8^r \lambda_t \qquad (20)$$
$$+ z_9^r \xi_t + z_{10}^r \upsilon_t$$

In solution (20), $z_0^r$ is the actual interest rate's steady-state equilibrium value.

Equation (20) underscores that the actual interest rate is subject to hysteresis (i.e., it does not depend exclusively on the current values of exogenous variables and shocks, but also on their respective past values). This is due to the fact that shocks to potential output, shocks to fiscal policy variables (taxes and public spending), and informative shocks are characterized by persistence effects, and therefore, private agents' consumption, saving, and investment decisions adjust progressively to their realizations. In other words, the demand and supply of capital change gradually when the economy is hit by fundamental shocks and/or the central bank and government release new information, which also implies a delay in the adjustment of the financial market specific price (the actual interest rate). However, the actual interest rate is also affected by idiosyncratic shocks and preference shocks, which, although lacking persistence, are relevant for private agents' consumption, saving, and investment choices.

*Proposition 2*

Along the common knowledge equilibrium path, the actual output $y_{t|t}^* = y_t^*$ is given by:

$$y_t^* = z_0^y + z_1^y \bar{y}_{t-2} + z_2^y \omega_{t-1} + z_3^y g_{t-1} + z_4^y \mathfrak{y}_t + z_5^y tax_{t-1} + z_6^y \mathfrak{Q}_t + z_7^y \chi_{t-1} + z_8^y \lambda_t \qquad (21)$$
$$+ z_9^y \xi_t + z_{10}^y \upsilon_t$$

where $z_0^y$ represents the actual output's steady-state value.

Actual GDP also shows some degree of inertia. Again, this is the natural consequence of the persistence of the potential output, fiscal, and news shocks, which implies that private agents' consumption, saving, and investment decisions adjust (reflecting in GDP) progressively when hit by these exogenous disturbances. Individual shocks and consumer preference shocks contribute equally to the instantaneous dynamics of equilibrium GDP.

Agents' expectations regarding future GDP are given by:

$$E_t[y_{t+1}^* | \mathcal{H}^t] = z_0^y + z_1^y \bar{y}_{t-1} + z_3^y g_t + z_5^y tax_t + z_7^y \chi_t \qquad (22)$$

Equation (22) indicates that, when formulating their expectations about the actual GDP at time $t + 1$, agents do not take into account those shocks that do not show persistence effects (i.e., consumer preference shocks and idiosyncratic shocks), as they do not imply significant and lasting deviations of GDP from its balanced growth path.

The forecasting error committed by agents regarding equilibrium GDP is a linear combination of white noise processes, i.e., it is a white noise process itself:

$$y_{t+1}^* - E_t[y_{t+1}^*|\mathcal{H}^t] = z_2^y \omega_t + z_4^y \mathfrak{y}_{t+1} + z_6^y \mathfrak{Q}_{t+1} + z_8^y \lambda_{t+1} + z_9^y \xi_{t+1} + z_{10}^y \upsilon_{t+1} \tag{23}$$

Therefore, consistently with the rational expectations hypothesis, the forecasting error associated with equilibrium GDP is a zero-mean process ($E_t[y_{t+1}^* - E_t[y_{t+1}^*|\mathcal{H}^t]|\mathcal{H}^t] = 0$). In other words, agents formulate forecasts regarding equilibrium GDP whose error is, on average, zero and non-systematic (i.e., non-persistent over time).

*Proposition 3*
Along the common knowledge equilibrium path, the output gap $\hat{y}_{t|t}^* = \hat{y}_t^*$ is given by:

$$\begin{aligned}
\hat{y}_t^* = {}& z_0^{\hat{y}} + z_1^{\hat{y}} \bar{y}_{t-2} + z_2^{\hat{y}} \omega_{t-1} + z_3^{\hat{y}} g_{t-1} + z_4^{\hat{y}} \mathfrak{y}_t + z_5^{\hat{y}} tax_{t-1} + z_6^{\hat{y}} \mathfrak{Q}_t + z_7^{\hat{y}} \chi_{t-1} + z_8^{\hat{y}} \lambda_t \\
& + z_9^{\hat{y}} \xi_t + z_{10}^{\hat{y}} \upsilon_t - \omega_t
\end{aligned} \tag{24}$$

where $z_0^{\hat{y}}$ is the output gap's steady-state equilibrium value.

The considerations that can be expressed regarding solution (24) are the same as those presented for the solution for equilibrium actual GDP: the process of output gap adjustment to informative shocks and to exogenous shocks affecting potential GDP, public spending, and tax revenue is gradual precisely because such disturbances lead to a progressive change in the economy's fundamentals. As can be seen from Table 3, the output gap's response to fiscal shocks is identical to that of actual GDP ($z_3^{\hat{y}} = z_3^y, z_4^{\hat{y}} = z_4^y, z_5^{\hat{y}} = z_5^y, z_6^{\hat{y}} = z_6^y$) and the same applies to the impact of the informative shocks ($z_7^{\hat{y}} = z_7^y, z_8^{\hat{y}} = z_8^y$), the demand shock ($z_9^{\hat{y}} = z_9^y$), and the idiosyncratic shock ($z_{10}^{\hat{y}} = z_{10}^y$). This is the natural consequence of the fact that the informative signal (including the unexpected informative shock at current time) and exogenous shocks to potential GDP, public spending, and taxes do not alter the economy's productive capacity (i.e., potential GDP), but rather act on actual GDP and its expectations. The output gap's response to a potential GDP shock, however, diverges from that of actual GDP because, by definition, $y_t = \bar{y}_t + \hat{y}_t$, where, according to equation (10), potential GDP $\bar{y}_t$ is directly affected by $\omega_t$ and its past realizations.

Agents' expectations regarding future output gap are given by

$$\begin{aligned}
E_t[\hat{y}_{t+1}^*|\mathcal{H}^t] = {}& z_0^{E\hat{y}} + z_1^{E\hat{y}} \bar{y}_{t-2} + z_2^{E\hat{y}} \omega_{t-1} + z_3^{E\hat{y}} g_{t-1} + z_4^{E\hat{y}} \mathfrak{y}_t + z_5^{E\hat{y}} tax_{t-1} + z_6^{E\hat{y}} \mathfrak{Q}_t \\
& + z_7^{E\hat{y}} \chi_{t-1} + z_8^{E\hat{y}} \lambda_t
\end{aligned} \tag{25}$$

that is, agents formulate their expectations about future GDP solely based on the informative signal they receive from public agents (including unexpected shocks) and those shocks that are characterized by persistence effects (i.e., those hitting the output gap, public spending, and tax revenues). This can be explained by the fact that these shocks are the only variables that, together with the informative signal, can lead to lasting and significant deviations of the output gap from its equilibrium path.

*Proposition 4*
Along the common knowledge equilibrium path, the expected inflation $E_t[\pi_{t+1|t}^*|\mathcal{H}^t] = E_t[\pi_{t+1}^*|\mathcal{H}^t]$ is given by:

$$E_t[\pi_{t+1}^* | \mathcal{H}^t] = z_0^{E\pi} + z_1^{E\pi}\bar{y}_{t-2} + z_2^{E\pi}\omega_{t-1} + z_3^{E\pi}g_{t-1} + z_4^{E\pi}\mathfrak{y}_t + z_5^{E\pi}tax_{t-1} + z_6^{E\pi}\mathfrak{Q}_t \quad (26)$$
$$+ z_7^{E\pi}\chi_{t-1} + z_8^{E\pi}\lambda_t + z_9^{E\pi}\xi_t + z_{10}^{E\pi}\upsilon_t + z_{11}^{E\pi}\omega_t + z_{12}^{E\pi}\epsilon_{t-1} + z_{13}^{E\pi}\varsigma_t$$

In equation (26), inflation expectations are determined by the informative signal coming from public agents (including unexpected shocks) and by shocks to potential GDP, public spending, and taxes. The reason for this inertial behavior of inflation expectations is that potential GDP (and therefore also the output gap and actual GDP), public spending, and tax revenues adjust gradually when hit by their respective shocks. Similarly, information disclosed by public agents is progressively incorporated by agents into their allocative decisions. In other words, since the economy's fundamentals adjust gradually to shocks, inflation expectations also adjust progressively.

*Proposition 5*
Along the common knowledge equilibrium path, the actual inflation rate $\pi_{t|t}^* = \pi_t^*$ is given by:

$$\pi_t^* = z_0^\pi + z_1^\pi\bar{y}_{t-2} + z_2^\pi\omega_{t-1} + z_3^\pi g_{t-1} + z_4^\pi\mathfrak{y}_t + z_5^\pi tax_{t-1} + z_6^\pi\mathfrak{Q}_t + z_7^\pi\chi_{t-1} + z_8^\pi\lambda_t \quad (27)$$
$$+ z_9^\pi\xi_t + z_{10}^\pi\upsilon_t + z_{11}^\pi\omega_t + z_{12}^\pi\epsilon_{t-1} + z_{13}^\pi\varsigma_t$$

where $z_0^\pi$ is the steady-state equilibrium value of the actual inflation rate.
In equation (27), the actual inflation rate exhibits inertial behavior which, once again, is explained by the persistence of the informative signal and shocks affecting potential GDP, fiscal policy variables (taxes and public spending), and inflation itself (i.e., the cost-push shock to inflation). As can be seen from the New Keynesian Phillips curve (equation (11)), the inertia of the inflation rate is induced through two transmission channels: the first consists of changes in inflation expectations, while the second is given by the output gap. Indeed, as highlighted in Table 6, in equation (27) the generic coefficient $z_i^\pi$ is a linear combination of the coefficients associated with the equilibrium solution for expected inflation and the output gap.

*Proposition 6*
Along the common knowledge equilibrium path, the household consumption $c_{t|t}^* = c_t^*$ is given by:

$$c_t^* = z_0^c + z_1^c\bar{y}_{t-2} + z_2^c\omega_{t-1} + z_3^c g_{t-1} + z_4^c\mathfrak{y}_t + z_5^c tax_{t-1} + z_6^c\mathfrak{Q}_t + z_7^c\chi_{t-1} + z_8^c\lambda_t \quad (28)$$
$$+ z_9^c\xi_t + z_{10}^c\upsilon_t$$

where $z_0^c$ is the household consumption's steady-state value.
Equation (28) indicates that the adjustment process of household consumption in response to the signal from public agents (including unexpected shocks) or to an exogenous shock affecting potential GDP and fiscal policy variables (taxes and public spending) is gradual, as households take time to update their expectations regarding the economy's fundamentals and thus modify their intertemporal consumption choices. Idiosyncratic shocks and preference shocks, on the other hand, as expected, have an immediate effect.

*Proposition 7*
Along the common knowledge equilibrium path, the firm's investment and (consequently) household saving $I_{t|t}^* = S_{t|t}^* = I_t^* = S_t^*$ are given by:

$$I_t^* = S_t^* = z_0^I + z_1^I\bar{y}_{t-2} + z_2^I\omega_{t-1} + z_3^I g_{t-1} + z_4^I\mathfrak{y}_t + z_5^I tax_{t-1} + z_6^I\mathfrak{Q}_t + z_7^I\chi_{t-1} + z_8^I\lambda_t \quad (29)$$
$$+ z_9^I\xi_t + z_{10}^I\upsilon_t$$

where the steady-state equilibrium value of the full information model for investment and saving is given by $z_0^I$.

The considerations for equation (29) are largely analogous to those for equation (28). The response of equilibrium investment and saving to the public agents' signal (including current informative shock) and to shocks affecting potential GDP, public spending, and taxes is gradual, as firms and households progressively update their expectations and (consequently) consumption and investment decisions. Conversely, idiosyncratic shocks and preference shocks have an immediate impact only, as they lack persistence.

*Proposition 8*
Along the common knowledge equilibrium path, the central bank's interest rate $i_{t|t}^* = i_t^*$ is given by:

$$i_t^* = z_0^i + z_1^i \bar{y}_{t-2} + z_2^i \omega_{t-1} + z_3^i g_{t-1} + z_4^i \mathfrak{y}_t + z_5^i tax_{t-1} + z_6^i \mathfrak{L}_t + z_7^i \chi_{t-1} + z_8^i \lambda_t \qquad (30)$$
$$+ z_9^i \xi_t + z_{10}^i \upsilon_t + z_{11}^i \omega_t + z_{12}^i \epsilon_{t-1} + z_{13}^i \varsigma_t$$

Equation (30) states that the central bank reacts to both current and past realizations of shocks affecting the output gap, inflation, and fiscal policy variables (public spending and taxes), as well as to the informative signal (including the unexpected informative shock at current time) coming from public agents. Since the effects of these shocks propagate gradually throughout the economy, the central bank's optimal reaction must be equally progressive. The change in the equilibrium policy rate following an idiosyncratic shock or a preference shock, however, is purely immediate, as these particular shocks have no persistence effects.

As shown in Table 9, the central bank's response to each shock strongly depends on the coefficients $\alpha_\pi$ and $\alpha_y$ of the Taylor rule (equation (13)), whose dimensions and signs reflect the central bank's own policy preferences. Indeed, the different combinations of $\alpha_\pi$ and $\alpha_y$ express the central bank's preferences for price stability and economic growth (and thus employment preservation).

However, it is important to note that equation (30) does not impose any a priori restrictions on the signs and sizes of the policy rate's responses to different types of shocks, regardless of whether the central bank chooses to commit itself to the Taylor principle ($\alpha_\pi > 1$) or not ($\alpha_\pi \leq 1$). In other words, even when it establishes price stability as its priority, the central bank might find it suboptimal to significantly increase the policy rate in response to an inflationary shock. This apparent contradiction is explained by the fact that the cost-push shock to inflation might show limited persistence and/or its effects might be offset by those of other exogenous shocks, rendering a decisive response unnecessary in the eyes of the monetary policy authority.

*Proposition 9*
Along the common knowledge equilibrium path, the actual unemployment rate $u_{t|t}^* = u_t^*$ is given by:

$$u_t^* = z_0^u + z_1^u \bar{y}_{t-2} + z_2^u \omega_{t-1} + z_3^u g_{t-1} + z_4^u \mathfrak{y}_t + z_5^u tax_{t-1} + z_6^u \mathfrak{L}_t + z_7^u \chi_{t-1} + z_8^u \lambda_t \qquad (31)$$
$$+ z_9^u \xi_t + z_{10}^u \upsilon_t + z_{11}^u \omega_t + z_{12}^u \bar{u}_{t-1} + \mathfrak{T}_t$$

where $z_0^u$ is the steady-state value of the unemployment rate that is determined by the real and nominal rigidities present in the economy.

Equation (30) clearly establishes that the unemployment rate is characterized by a hysteresis effect, which, once again, is explained by the persistence of the informative signal coming from fully informed public agents (which includes unexpected informative shocks at current time), shocks affecting potential GDP and fiscal policy variables (taxes and public spending), and the persistence in the natural unemployment rate. Instantaneous changes in the interest rate, on the other hand, are determined by idiosyncratic and preference shocks.

In equilibrium, the expected value of the unemployment rate at time $t + 1$ is given by:

$$E_t[u_{t+1}^* | \mathcal{H}^t] = z_0^{Eu} + z_1^{Eu}\bar{y}_{t-2} + z_2^{Eu}\omega_{t-1} + z_3^{Eu}g_{t-1} + z_4^{Eu}\mathfrak{y}_t + z_5^{Eu}tax_{t-1} + z_6^{Eu}\mathfrak{L}_t \quad (32)$$
$$+ z_7^{Eu}\chi_{t-1} + z_8^{Eu}\lambda_t + z_9^{Eu}\bar{u}_{t-1} + z_{10}^{Eu}\mathfrak{T}_t$$

Equation (32) represents households' expectation of being unemployed at time $t+1$. Once again, this expectation is a linear combination of the informative signal (including unexpected informational shocks in the current period) and shocks characterized by persistence effects (i.e., those to potential GDP, government spending, taxes, and the natural unemployment rate).

The representative household's job insecurity at time $t+1$, or rather, the perceived risk of no longer being employed at time $t+1$ for those who are not unemployed in the current period, is given by the following expression:

$$E_t[u_{t+1}^* - u_0 | \mathcal{H}^t]$$
$$= z_1^{Eu}\bar{y}_{t-2} + z_2^{Eu}\omega_{t-1} + z_3^{Eu}g_{t-1} + z_4^{Eu}\mathfrak{y}_t + z_5^{Eu}tax_{t-1} + z_6^{Eu}\mathfrak{L}_t \quad (33)$$
$$+ z_7^{Eu}\chi_{t-1} + z_8^{Eu}\lambda_t + z_9^{Eu}\bar{u}_{t-1} + z_{10}^{Eu}\mathfrak{T}_t$$

that is the expected deviation of actual unemployment rate from its steady-state value between time $t$ and time $t+1$.

The intuition behind this result is that households' job insecurity is determined by the set of exogenous shocks affecting the economy in the current period, causing a temporary deviation of the actual unemployment rate from its steady-state level ($z_0^u$).

Equation (33) indicates that the persistence of the signal released by public agents (including unexpected informational shocks) and shocks to potential GDP, the natural unemployment rate, taxes, and public spending cause job insecurity to adjust gradually over time (i.e., equilibrium job insecurity is subject to hysteresis). It is very interesting to note that job insecurity is not affected by idiosyncratic shocks or preference shocks. This is probably due to the fact that, not being persistent, individual-level shocks do not generate changes in expectations regarding the economy's long-term fundamentals, and thus, households, at the aggregate level, do not take them into account when forming their expectations about the prosecution of their own job career.

On the other hand, from the perspective of the representative firm, equation (33) expresses the decision to hire or fire workers in the near future on the basis of their current information set. This interpretation is in line with the established theoretical and empirical literature according to which companies choose today the working hours they need to achieve the levels of output desired in the near future upon forming their optimal expectations about the future state of the economy (Lucchese & Pianta, 2012; Vota, 2022).

In the common knowledge solution of this model, all endogenous variables move along a saddle path, namely, their long-run convergence towards their respective steady-state equilibria (which are generically labeled as $z_0^j$ in equations (20)-(33) and consist of nonlinear combinations of the parameters of the model's structural form) is the result of both exogenous shocks that propagate instantaneously (i.e., preference shocks and idiosyncratic shocks) and shocks that exhibit short-run persistence.

Shocks that follow a stationary AR(1) process cannot cause significantly lasting deviations of endogenous variables from their respective saddle paths because, by assumption, they are stationary. Meanwhile, (at least in principle) white noise shocks can generate substantial deviations that lead to explosive dynamics (though this is a rare eventuality that can only be triggered in the rare cases where the magnitude of the white noise shock is such that it compromises the economic system's absorption capacity).

It is important to highlight that common knowledge solutions are not achieved by imposing ad hoc restrictions on the signs of the coefficients in the model's reduced-form, and, of course, this is true also for the parameters associated with the informative signal $\chi_t$ (which includes the unexpected informative shocks at the current time $\lambda_t$). This means that the model's common knowledge solution reflects the "Paradox of Transparency", according to which nothing guarantees that a clear and transparent communication strategy from public agents (i.e., one without noise) necessarily leads to welfare improvement.

This result stems from the fact that unexpected news shocks ($\lambda_t$) can either improve or worsen private agents' expectations regarding the economy's fundamentals. This has significant (and potentially welfare-detrimental) consequences for investment, consumption, saving, production activities, and unemployment.

The Paradox of Transparency also impacts job insecurity: the public sector's diffusion activities of otherwise unknown information influence private agents' long-run expectations about the economy's fundamentals, and therefore, workers' expectations about the continuity of their own careers.

## 5. Asymmetric information analysis

To find the solution of the model with asymmetric information, it is first necessary to write the conditioned-down system (i.e., the expression used to map the relationship between observables and unobservables and derive the projection conditions), bearing in mind that now the informative signal sent by the public agents to the private ones is affected by a noisy term ($\Psi_{t|t} = \chi_{t|t} + \Xi_{t|t}$).

In the system of equations below, each model's endogenous variables is expressed as a linear function of the exogenous ones:

$$r_t^* = z_0^r + z_1^r \bar{y}_{t-2} + z_2^r \omega_{t-1|t} + z_3^r g_{t-1|t} + z_4^r \mathfrak{y}_{t|t} + z_5^r tax_{t-1|t} + z_6^r \mathfrak{L}_{t|t} + z_7^r \chi_{t-1} \tag{34.1}$$
$$+ z_8^r \lambda_t + z_9^r \xi_{t|t} + z_{10}^r \upsilon_{t|t}$$

$$y_{t|t} = z_0^y + z_1^y \bar{y}_{t-2} + z_2^y \omega_{t-1|t} + z_3^y g_{t-1|t} + z_4^y \mathfrak{y}_{t|t} + z_5^y tax_{t-1|t} + z_6^y \mathfrak{L}_{t|t} + z_7^y \chi_{t-1} \tag{34.2}$$
$$+ z_8^y \lambda_t + z_9^y \xi_{t|t} + z_{10}^y \upsilon_{t|t}$$

$$\hat{y}_{t|t}^* = z_0^{\hat{y}} + z_1^{\hat{y}} \bar{y}_{t-2} + z_2^{\hat{y}} \omega_{t-1|t} + z_3^{\hat{y}} g_{t-1|t} + z_4^{\hat{y}} \mathfrak{y}_{t|t} + z_5^{\hat{y}} tax_{t-1|t} + z_6^{\hat{y}} \mathfrak{L}_{t|t} + z_7^{\hat{y}} \chi_{t-1} \tag{34.3}$$
$$+ z_8^{\hat{y}} \lambda_t + z_9^{\hat{y}} \xi_{t|t} + z_{10}^{\hat{y}} \upsilon_{t|t} - \omega_{t|t}$$

$$\pi_t^* = z_0^\pi + z_1^\pi \bar{y}_{t-2} + z_2^\pi \omega_{t-1|t} + z_3^\pi g_{t-1|t} + z_4^\pi \mathfrak{y}_{t|t} + z_5^\pi tax_{t-1|t} + z_6^\pi \mathfrak{L}_{t|t} + z_7^\pi \chi_{t-1} \tag{34.4}$$
$$+ z_8^\pi \lambda_t + z_9^\pi \xi_{t|t} + z_{10}^\pi \upsilon_{t|t} + z_{11}^\pi \omega_{t|t} + z_{12}^\pi \epsilon_{t-1|t} + z_{13}^\pi \varsigma_{t|t}$$

$$c_{t|t}^* = z_0^c + z_1^c \bar{y}_{t-2} + z_2^c \omega_{t-1|t} + z_3^c g_{t-1|t} + z_4^c \mathfrak{y}_{t|t} + z_5^c tax_{t-1|t} + z_6^c \mathfrak{L}_{t|t} + z_7^c \chi_{t-1} \tag{34.5}$$
$$+ z_8^c \lambda_t + z_9^c \xi_{t|t} + z_{10}^c \upsilon_{t|t}$$

$$I_{t|t}^* = S_{t|t}^* = z_0^I + z_1^I \bar{y}_{t-2} + z_2^I \omega_{t-1|t} + z_3^I g_{t-1|t} + z_4^I \mathfrak{y}_{t|t} + z_5^I tax_{t-1|t} + z_6^I \mathfrak{L}_{t|t} \tag{34.6}$$
$$+ z_7^I \chi_{t-1} + z_8^I \lambda_t + z_9^I \xi_{t|t} + z_{10}^I \upsilon_{t|t}$$

$$i_{t|t}^* = z_0^i + z_1^i \bar{y}_{t-2} + z_2^i \omega_{t-1|t} + z_3^i g_{t-1|t} + z_4^i \mathfrak{y}_{t|t} + z_5^i tax_{t-1|t} + z_6^i \mathfrak{L}_{t|t} + z_7^i \chi_{t-1} \tag{34.7}$$
$$+ z_8^i \lambda_t + z_9^i \xi_{t|t} + z_{10}^i \upsilon_{t|t} + z_{11}^i \omega_{t|t} + z_{12}^i \epsilon_{t-1|t} + z_{13}^i \varsigma_{t|t}$$

$$u_{t|t}^* = z_0^u + z_1^u \bar{y}_{t-2} + z_2^u \omega_{t-1|t} + z_3^u g_{t-1|t} + z_4^u \mathfrak{y}_{t|t} + z_5^u tax_{t-1|t} + z_6^u \mathfrak{L}_{t|t} + z_7^u \chi_{t-1} \tag{34.8}$$
$$+ z_8^u \lambda_t + z_9^u \xi_{t|t} + z_{10}^u \upsilon_{t|t} + z_{11}^u \omega_{t|t} + z_{12}^u \bar{u}_{t-1|t} + \mathfrak{T}_{t|t}$$

$$\Psi_{t|t} = z_1^\psi \chi_{t-1} + \lambda_t + \Xi_{t|t} \tag{34.9}$$

where $z_1^{\psi} = \rho_{\chi}$.

Note that the conditioned-down system is now not isomorphic to the full information solution, both due to the presence of equation (34.9) and because the expression for $\chi_t$ (and, consequently, $\chi_{t-1}$) used in equations (34.1)-(34.8) differs from that of the full information solution. Indeed, in the common knowledge setting $\chi_t = \Psi_{t|t} = \Psi_t$, while in the case of asymmetric information $\chi_t = \Psi_{t|t} - \Xi_{t|t}$ where $\Psi_{t|t}$ is an additional variable for which households and firms must form an optimal projection.

Furthermore, the system formed by equations (34.1)-(34.9) admits infinite solutions for the model's observables $(r_t^*, \bar{y}_{t-2}, \chi_{t-1}, \lambda_t)$, because it consists of nine equations in nineteen unknowns $(y_{t|t}^*, \hat{y}_{t|t}^*, \pi_{t|t}^*, c_{t|t}^*, I_{t|t}^*, i_{t|t}^*, u_{t|t}^*, \omega_{t-1|t}, g_{t-1|t}, \mathfrak{y}_{t|t}, tax_{t-1|t}, \mathfrak{L}_{t|t}, \xi_{t|t}, \upsilon_{t|t}, \omega_{t|t}, \bar{u}_{t-1|t}, \epsilon_{t-1|t}, \varsigma_{t|t}, \mathfrak{T}_{t|t})$. However, it cannot be ruled out a priori that, despite the infinite mathematical solutions of the conditioned-down system, the application of the "projection conditions" derived from rational expectations and the imposition of other economically significant stability conditions (such as transversality conditions) can select a unique equilibrium path (or at least limit the solutions in ways not immediately evident from a simple equation/unknown count).

Precisely for this reason, it is fundamental to apply the Blanchard-Kahn method. This method does not merely count equations and unknowns, but analyzes the dynamic structure of the system of stochastic differential equations, examining the eigenvalues of the transition matrix and comparing them with the number of non-predetermined variables. Only in this way can it be determined with certainty whether there is a unique stable solution, infinitely many stable solutions (economic indeterminacy), or no stable solution.

The Blanchard-Kahn method requires the model to be rewritten in its first-order state-space form:

$$E_t\left[x_{t+1}|\mathcal{H}^t\right] = Ax_t + Bw_t \tag{35}$$

where:
- $x_t$ is the vector of predetermined state variables:
- $A$ is the state-transition matrix;
- $w_t$ is the exogenous shock matrix;
- $I$ is the identity matrix;
- $B$ is the shock response matrix.

and that, subsequently, the eigenvalues of the matrix $A$ are calculated and counted. According to Blanchard and Kahn, if the number of stable eigenvalues (i.e., those whose modulus is strictly less than one) equals the number of predetermined variables, then the system is determined and the model admits only one solution (equilibrium determinacy). If the number of stable eigenvalues is greater than the number of predetermined variables, there is equilibrium indeterminacy. Finally, if the number of stable eigenvalues is less than the number of predetermined variables, the model admits no equilibrium.

Exploiting the AR(1) property of variables $\bar{y}_t, g_t, tax_t, \chi_t$, and $\epsilon_t$, it is possible to obtain the following first-order state space representation of the model with asymmetric information:

$$
\begin{aligned}
E_t[r_{t+1}^*|\mathcal{H}^t] = {} & z_0^r + z_1^r \rho_y^3 \bar{y}_{t-4} + z_1^r \rho_y^2 \omega_{t-3} + z_1^r \rho_y \bar{y}_{t-2} - z_1^r \rho_y^3 \bar{y}_{t-3} + z_1^r \omega_{t-1} \\
& + z_3^r \rho_g^4 g_{t-4} + z_3^r \rho_g^3 \mathfrak{y}_{t-3} + z_3^r \rho_g^2 g_{t-2} - z_3^r \rho_g^3 g_{t-3} + \rho_g z_3^r \mathfrak{y}_{t-1} + z_3^r \mathfrak{y}_t \\
& + \rho_{tax} z_5^r tax_{t-1} + z_5^r \mathfrak{L}_t + \rho_{\chi} z_7^r \chi_{t-1} + z_7^r \lambda_t
\end{aligned} \tag{36.1}
$$

$$
\begin{aligned}
E_t[y_{t+1}^*|\mathcal{H}^t] = {} & z_0^y + z_1^y \rho_y^3 \bar{y}_{t-4} + z_1^y \rho_y^2 \omega_{t-3} + z_1^y \rho_y \bar{y}_{t-2} - z_1^y \rho_y^2 \bar{y}_{t-3} + z_1^y \omega_{t-1} \\
& + z_3^y \rho_g^4 g_{t-4} + z_3^y \rho_g^3 \mathfrak{y}_{t-3} + z_3^y \rho_g^2 g_{t-2} - z_3^y \rho_g^3 g_{t-3} + \rho_g z_3^y \mathfrak{y}_{t-1} + z_3^y \mathfrak{y}_t \\
& + \rho_{tax} z_5^y tax_{t-1} + z_5^y \mathfrak{L}_t + \rho_{\chi} z_7^y \chi_{t-1} + z_7^y \lambda_t
\end{aligned} \tag{36.2}
$$

$$E_t[\hat{y}^*_{t+1}|\mathcal{H}^t] = z_0^{\hat{y}} + z_1^{\hat{y}}\rho_y^3\bar{y}_{t-4} + z_1^{\hat{y}}\rho_y^2\omega_{t-3} + z_1^{\hat{y}}\rho_y\bar{y}_{t-2} - z_1^{\hat{y}}\rho_y^2\bar{y}_{t-3} + z_1^{\hat{y}}\omega_{t-1}$$
$$+ z_3^{\hat{y}}\rho_g^4 g_{t-4} + z_3^{\hat{y}}\rho_g^3\mathfrak{y}_{t-3} + z_3^{\hat{y}}\rho_g^2 g_{t-2} - z_3^{\hat{y}}\rho_g^3 g_{t-3} + \rho_g z_3^{\hat{y}}\mathfrak{y}_{t-1} + z_3^{\hat{y}}\mathfrak{y}_t$$
$$+ \rho_{tax}z_5^{\hat{y}}tax_{t-1} + z_5^{\hat{y}}\mathfrak{L}_t + \rho_\chi z_7^{\hat{y}}\chi_{t-1} + z_7^{\hat{y}}\lambda_t \qquad (36.3)$$

$$E_t[\pi^*_{t+1}|\mathcal{H}^t] = z_0^{\pi} + z_1^{\pi}\rho_y^3\bar{y}_{t-4} + z_1^{\pi}\rho_y^2\omega_{t-3} + z_1^{\pi}\rho_y\bar{y}_{t-2} - z_1^{\pi}\rho_y^2\bar{y}_{t-3} + z_1^{\pi}\omega_{t-1}$$
$$+ z_3^{\pi}\rho_g^4 g_{t-4} + z_3^{\pi}\rho_g^3\mathfrak{y}_{t-3} + z_3^{\pi}\rho_g^2 g_{t-2} - z_3^{\pi}\rho_g^3 g_{t-3} + \rho_g z_3^{\pi}\mathfrak{y}_{t-1} + z_3^{\pi}\mathfrak{y}_t$$
$$+ \rho_{tax}z_5^{\pi}tax_{t-1} + z_5^{\pi}\mathfrak{L}_t + \rho_\chi z_7^{\pi}\chi_{t-1} + z_7^{\pi}\lambda_t + \rho_\epsilon z_{12}^{\pi}\epsilon_{t-1} + z_{12}^{\pi}\varsigma_t \qquad (36.4)$$

$$E_t[c^*_{t+1}|\mathcal{H}^t] = z_0^{c} + z_1^{c}\rho_y^3\bar{y}_{t-4} + z_1^{c}\rho_y^2\omega_{t-3} + z_1^{c}\rho_y\bar{y}_{t-2} - z_1^{c}\rho_y^2\bar{y}_{t-3} + z_1^{c}\omega_{t-1}$$
$$+ z_3^{c}\rho_g^4 g_{t-4} + z_3^{c}\rho_g^3\mathfrak{y}_{t-3} + z_3^{c}\rho_g^2 g_{t-2} - z_3^{c}\rho_g^3 g_{t-3} + \rho_g z_3^{c}\mathfrak{y}_{t-1} + z_3^{c}\mathfrak{y}_t$$
$$+ \rho_{tax}z_5^{c}tax_{t-1} + z_5^{c}\mathfrak{L}_t + \rho_\chi z_7^{c}\chi_{t-1} + z_7^{c}\lambda_t \qquad (36.5)$$

$$E_t[I^*_{t+1}|\mathcal{H}^t] = z_0^{I} + z_1^{I}\rho_y^3\bar{y}_{t-4} + z_1^{I}\rho_y^2\omega_{t-3} + z_1^{I}\rho_y\bar{y}_{t-2} - z_1^{I}\rho_y^2\bar{y}_{t-3} + z_1^{I}\omega_{t-1}$$
$$+ z_3^{I}\rho_g^4 g_{t-4} + z_3^{I}\rho_g^3\mathfrak{y}_{t-3} + z_3^{I}\rho_g^2 g_{t-2} - z_3^{I}\rho_g^3 g_{t-3} + \rho_g z_3^{I}\mathfrak{y}_{t-1} + z_3^{I}\mathfrak{y}_t$$
$$+ \rho_{tax}z_5^{I}tax_{t-1} + z_5^{I}\mathfrak{L}_t + \rho_\chi z_7^{I}\chi_{t-1} + z_7^{I}\lambda_t \qquad (36.6)$$

$$E_t[i^*_{t+1}|\mathcal{H}^t] = z_0^{i} + z_1^{i}\rho_y^3\bar{y}_{t-4} + z_1^{i}\rho_y^2\omega_{t-3} + z_1^{i}\rho_y\bar{y}_{t-2} - z_1^{i}\rho_y^2\bar{y}_{t-3} + z_1^{i}\omega_{t-1}$$
$$+ z_3^{i}\rho_g^4 g_{t-4} + z_3^{i}\rho_g^3\mathfrak{y}_{t-3} + z_3^{i}\rho_g^2 g_{t-2} - z_3^{i}\rho_g^3 g_{t-3} + \rho_g z_3^{i}\mathfrak{y}_{t-1} + z_3^{i}\mathfrak{y}_t$$
$$+ \rho_{tax}z_5^{i}tax_{t-1} + z_5^{i}\mathfrak{L}_t + \rho_\chi z_7^{i}\chi_{t-1} + z_7^{i}\lambda_t + \rho_\epsilon z_{12}^{i}\epsilon_{t-1} + z_{12}^{i}\varsigma_t \qquad (36.7)$$

$$E_t[u^*_{t+1}|\mathcal{H}^t] = z_0^{u} + z_1^{u}\rho_y^3\bar{y}_{t-4} + z_1^{u}\rho_y^2\omega_{t-3} + z_1^{u}\rho_y\bar{y}_{t-2} - z_1^{u}\rho_y^2\bar{y}_{t-3} + z_1^{u}\omega_{t-1}$$
$$+ z_3^{u}\rho_g^4 g_{t-4} + z_3^{u}\rho_g^3\mathfrak{y}_{t-3} + z_3^{u}\rho_g^2 g_{t-2} - z_3^{u}\rho_g^3 g_{t-3} + \rho_g z_3^{u}\mathfrak{y}_{t-1} + z_3^{u}\mathfrak{y}_t$$
$$+ \rho_{tax}z_5^{u}tax_{t-1} + z_5^{u}\mathfrak{L}_t + \rho_\chi z_7^{u}\chi_{t-1} + z_7^{u}\lambda_t \qquad (36.8)$$

$$E_t[\Psi^*_{t+1}|\mathcal{H}^t] = \rho_\chi^2\chi_{t-1} + \lambda_t \qquad (36.9)$$

whose state-transition matrix is a square matrix of order nine.
The characteristic equation associated with this state-space representation is:

$$det\begin{bmatrix} z_1^r\rho_y^3 - a & z_1^r\rho_y & -z_1^r\rho_y^2 & z_3^r\rho_g^4 & z_3^r\rho_g^2 & -z_3^r\rho_g^3 & \rho_{tax}z_5^r & \rho_\chi z_7^r & 0 \\ z_1^y\rho_y^3 & z_1^y\rho_y - a & -z_1^y\rho_y^2 & z_3^y\rho_g^4 & z_3^y\rho_g^2 & -z_3^y\rho_g^3 & \rho_{tax}z_5^y & \rho_\chi z_7^y & 0 \\ z_1^{\hat{y}}\rho_y^3 & z_1^{\hat{y}}\rho_y & -z_1^{\hat{y}}\rho_y^2 - a & z_3^{\hat{y}}\rho_g^4 & z_3^{\hat{y}}\rho_g^2 & -z_3^{\hat{y}}\rho_g^3 & \rho_{tax}z_5^{\hat{y}} & \rho_\chi z_7^{\hat{y}} & 0 \\ z_1^{\pi}\rho_y^3 & z_1^{\pi}\rho_y & -z_1^{\pi}\rho_y^2 & z_3^{\pi}\rho_g^4 - a & z_3^{\pi}\rho_g^2 & -z_3^{\pi}\rho_g^3 & \rho_{tax}z_5^{\pi} & \rho_\chi z_7^{\pi} & \rho_\epsilon z_{12}^{\pi} \\ z_1^{c}\rho_y^3 & z_1^{c}\rho_y & -z_1^{c}\rho_y^2 & z_3^{c}\rho_g^4 & z_3^{c}\rho_g^2 - a & -z_3^{c}\rho_g^3 & \rho_{tax}z_5^{c} & \rho_\chi z_7^{c} & 0 \\ z_1^{I}\rho_y^3 & z_1^{I}\rho_y & -z_1^{I}\rho_y^2 & z_3^{I}\rho_g^4 & z_3^{I}\rho_g^2 & -z_3^{I}\rho_g^3 - a & \rho_{tax}z_5^{I} & \rho_\chi z_7^{I} & 0 \\ z_1^{i}\rho_y^3 & z_1^{i}\rho_y^2 & -z_1^{i}\rho_y^2 & z_3^{i}\rho_g^4 & z_3^{i}\rho_g^2 & -z_3^{i}\rho_g^3 & \rho_{tax}z_5^{i} - a & \rho_\chi z_7^{i} & \rho_\epsilon z_{12}^{i} \\ z_1^{u}\rho_y^3 & z_1^{u}\rho_y & -z_1^{u}\rho_y^2 & z_3^{u}\rho_g^4 & z_3^{u}\rho_g^2 & -z_3^{u}\rho_g^3 & \rho_{tax}z_5^{u} & \rho_\chi z_7^{u} - a & 0 \\ 0 & 0 & 0 & 0 & 0 & 0 & 0 & \rho_\chi^2 & -a \end{bmatrix} = 0 \quad (37)$$

where $a$ is the generic eigenvalue.
Equation (37) leads to a ninth-degree polynomial in $a$ whose generic coefficients $k_i$ with $i = 1, ..., 9$ are nonlinear combinations of the parameters of the state-space representation (36.1)-(36.9):

$$k_0 + k_1 a + k_2 a^2 + k_3 a^3 + k_4 a^4 + k_5 a^5 + k_6 a^6 + k_7 a^7 + k_8 a^8 + k_9 a^9 = 0 \qquad (38)$$

Due to its complexity, equation (38) is analytically intractable without specifying the punctual values of the generic coefficient $z_i^j$ in equation (37) and deriving the corresponding value of the generic coefficient $k_i$ in equation (38).

Although the problem of equilibrium determinacy cannot be solved analytically (and, consequently, the possible equilibria of the model with asymmetric information cannot be characterized), it is possible to develop some considerations regarding the contribution of job insecurity to equilibrium determination.

In fact, note that the state-space representation can be re-expressed in terms of the deviation of state variables from their respective full information steady-state values:

$$E_t[r_{t+1}^* - z_0^r | \mathcal{H}^t] \qquad (39.1)$$
$$\begin{aligned} = & z_1^r \rho_y^3 \bar{y}_{t-4} + z_1^r \rho_y^2 \omega_{t-3} + z_1^r \rho_y \bar{y}_{t-2} - z_1^r \rho_y^2 \bar{y}_{t-3} + z_1^r \omega_{t-1} \\ & + z_3^r \rho_g^4 g_{t-4} + z_3^r \rho_g^3 \mathfrak{y}_{t-3} + z_3^r \rho_g^2 g_{t-2} - z_3^r \rho_g^3 g_{t-3} + \rho_g z_3^r \mathfrak{y}_{t-1} + z_3^r \mathfrak{y}_t \\ & + \rho_{tax} z_5^r tax_{t-1} + z_5^r \mathfrak{L}_t + \rho_\chi z_7^r \chi_{t-1} + z_7^r \lambda_t \end{aligned}$$

$$E_t[y_{t+1}^* - z_0^y | \mathcal{H}^t] \qquad (39.2)$$
$$\begin{aligned} = & z_1^y \rho_y^3 \bar{y}_{t-4} + z_1^y \rho_y^2 \omega_{t-3} + z_1^y \rho_y \bar{y}_{t-2} - z_1^y \rho_y^2 \bar{y}_{t-3} + z_1^y \omega_{t-1} \\ & + z_3^y \rho_g^4 g_{t-4} + z_3^y \rho_g^3 \mathfrak{y}_{t-3} + z_3^y \rho_g^2 g_{t-2} - z_3^y \rho_g^3 g_{t-3} + \rho_g z_3^y \mathfrak{y}_{t-1} + z_3^y \mathfrak{y}_t \\ & + \rho_{tax} z_5^y tax_{t-1} + z_5^y \mathfrak{L}_t + \rho_\chi z_7^y \chi_{t-1} + z_7^y \lambda_t \end{aligned}$$

$$E_t[\hat{y}_{t+1}^* - z_0^{\hat{y}} | \mathcal{H}^t] \qquad (39.3)$$
$$\begin{aligned} = & z_1^{\hat{y}} \rho_y^3 \bar{y}_{t-4} + z_1^{\hat{y}} \rho_y^2 \omega_{t-3} + z_1^{\hat{y}} \rho_y \bar{y}_{t-2} - z_1^{\hat{y}} \rho_y^2 \bar{y}_{t-3} + z_1^{\hat{y}} \omega_{t-1} \\ & + z_3^{\hat{y}} \rho_g^4 g_{t-4} + z_3^{\hat{y}} \rho_g^3 \mathfrak{y}_{t-3} + z_3^{\hat{y}} \rho_g^2 g_{t-2} - z_3^{\hat{y}} \rho_g^3 g_{t-3} + \rho_g z_3^{\hat{y}} \mathfrak{y}_{t-1} + z_3^{\hat{y}} \mathfrak{y}_t \\ & + \rho_{tax} z_5^{\hat{y}} tax_{t-1} + z_5^{\hat{y}} \mathfrak{L}_t + \rho_\chi z_7^{\hat{y}} \chi_{t-1} + z_7^{\hat{y}} \lambda_t \end{aligned}$$

$$E_t[\pi_{t+1}^* - z_0^\pi | \mathcal{H}^t] \qquad (39.4)$$
$$\begin{aligned} = & z_1^\pi \rho_y^3 \bar{y}_{t-4} + z_1^\pi \rho_y^2 \omega_{t-3} + z_1^\pi \rho_y \bar{y}_{t-2} - z_1^\pi \rho_y^2 \bar{y}_{t-3} + z_1^\pi \omega_{t-1} \\ & + z_3^\pi \rho_g^4 g_{t-4} + z_3^\pi \rho_g^3 \mathfrak{y}_{t-3} + z_3^\pi \rho_g^2 g_{t-2} - z_3^\pi \rho_g^3 g_{t-3} + \rho_g z_3^\pi \mathfrak{y}_{t-1} + z_3^\pi \mathfrak{y}_t \\ & + \rho_{tax} z_5^\pi tax_{t-1} + z_5^\pi \mathfrak{L}_t + \rho_\chi z_7^\pi \chi_{t-1} + z_7^\pi \lambda_t + \rho_\epsilon z_{12}^\pi \epsilon_{t-1} + z_{12}^\pi \varsigma_t \end{aligned}$$

$$E_t[c_{t+1}^* - z_0^c | \mathcal{H}^t] \qquad (39.5)$$
$$\begin{aligned} = & z_1^c \rho_y^3 \bar{y}_{t-4} + z_1^c \rho_y^2 \omega_{t-3} + z_1^c \rho_y \bar{y}_{t-2} - z_1^c \rho_y^2 \bar{y}_{t-3} + z_1^c \omega_{t-1} \\ & + z_3^c \rho_g^4 g_{t-4} + z_3^c \rho_g^3 \mathfrak{y}_{t-3} + z_3^c \rho_g^2 g_{t-2} - z_3^c \rho_g^3 g_{t-3} + \rho_g z_3^c \mathfrak{y}_{t-1} + z_3^c \mathfrak{y}_t \\ & + \rho_{tax} z_5^c tax_{t-1} + z_5^c \mathfrak{L}_t + \rho_\chi z_7^c \chi_{t-1} + z_7^c \lambda_t \end{aligned}$$

$$E_t[I_{t+1}^* - z_0^I | \mathcal{H}^t] \qquad (39.6)$$
$$\begin{aligned} = & z_1^I \rho_y^3 \bar{y}_{t-4} + z_1^I \rho_y^2 \omega_{t-3} + z_1^I \rho_y \bar{y}_{t-2} - z_1^I \rho_y^2 \bar{y}_{t-3} + z_1^I \omega_{t-1} + z_3^I \rho_g^4 g_{t-4} \\ & + z_3^I \rho_g^3 \mathfrak{y}_{t-3} + z_3^I \rho_g^2 g_{t-2} - z_3^I \rho_g^3 g_{t-3} + \rho_g z_3^I \mathfrak{y}_{t-1} + z_3^I \mathfrak{y}_t \\ & + \rho_{tax} z_5^I tax_{t-1} + z_5^I \mathfrak{L}_t + \rho_\chi z_7^I \chi_{t-1} + z_7^I \lambda_t \end{aligned}$$

$$E_t[i_{t+1}^* - z_0^i | \mathcal{H}^t] \qquad (39.7)$$
$$\begin{aligned} = & z_1^i \rho_y^3 \bar{y}_{t-4} + z_1^i \rho_y^2 \omega_{t-3} + z_1^i \rho_y \bar{y}_{t-2} - z_1^i \rho_y^2 \bar{y}_{t-3} + z_1^i \omega_{t-1} + z_3^i \rho_g^4 g_{t-4} \\ & + z_3^i \rho_g^3 \mathfrak{y}_{t-3} + z_3^i \rho_g^2 g_{t-2} - z_3^i \rho_g^3 g_{t-3} + \rho_g z_3^i \mathfrak{y}_{t-1} + z_3^i \mathfrak{y}_t \\ & + \rho_{tax} z_5^i tax_{t-1} + z_5^i \mathfrak{L}_t + \rho_\chi z_7^i \chi_{t-1} + z_7^i \lambda_t + \rho_\epsilon z_{12}^i \epsilon_{t-1} + z_{12}^i \varsigma_t \end{aligned}$$

$$E_t[u_{t+1}^* - z_0^u | \mathcal{H}^t] \tag{39.8}$$
$$= z_1^u \rho_y^3 \bar{y}_{t-4} + z_1^u \rho_y^2 \omega_{t-3} + z_1^u \rho_y \bar{y}_{t-2} - z_1^u \rho_y^3 \bar{y}_{t-3} + z_1^u \omega_{t-1}$$
$$+ z_3^u \rho_g^4 g_{t-4} + z_1^u \rho_g^3 \mathfrak{y}_{t-3} + z_3^u \rho_g^2 g_{t-2} - z_3^u \rho_g^3 g_{t-3} + \rho_g z_3^u \mathfrak{y}_{t-1} + z_3^u \mathfrak{y}_t$$
$$+ \rho_{tax} z_5^u tax_{t-1} + z_5^u \mathfrak{L}_t + \rho_\chi z_7^u \chi_{t-1} + z_7^u \lambda_t$$

$$E_t[\Psi_{t+1}^* | \mathcal{H}^t] = \rho_\chi^2 \chi_{t-1} + \lambda_t \tag{39.9}$$

where, once again, equation (39.8) is job insecurity.

Naturally, since the characteristic equation associated with the representation (39.1)-(39.9) of the model with asymmetric information is perfectly equivalent to that of the previous representation (36.1)-(36.9), the two systems share the same characteristic polynomial (equation (38)), and therefore also the same set of theoretical solutions.

This means that the mechanism through which expectations about the model's state variables are formed under asymmetric information alters the structure of the dynamic equations (as highlighted by the state-space representation (39.1)-(39.9)), making job insecurity an active element in defining the system's eigenvalues through the elements of matrix $A$.

In fact, by definition of eigenvalue, it holds:

$$Av = av \tag{40}$$

where $v$ s the eigenvector associated with the eigenvalues comprised in $a$.

Equation (40) establishes the relationship between the coefficients of the state-space representation equations (39.1)-(39.9) and the eigenvalues and indicates that expectations regarding the deviations of state variables from their respective steady-state values (thus including job insecurity) jointly define the existence and uniqueness of the equilibrium in the model with asymmetric information.

In other words, in the case of asymmetric information and noisy communication by public agents, it is impossible to establish with certainty whether at least one equilibrium exists within the economy without having specified the exact values of the structural form parameters of the model. However, it can be stated with certainty that the existence and uniqueness of macroeconomic equilibrium are jointly determined by agents' expectations regarding state variables. Expectations about the future unemployment rate, and consequently job insecurity, contribute to the determination process in a potentially decisive way.

## 6. Discussion

The model presented in this work is close in spirit to Han (2024), who analyses the problem of expectations misalignment in a New Keynesian setting in which private agents formulate their optimal estimates (projections) of the unobservable endogenous variables on the basis of a noisy signal about the exogenous shocks, while the central bank acquires information over time through an adaptive learning process. In his study, Han (2024) does not address the equilibrium determinacy problem but clearly documents that, because of the information asymmetry and noise, non-professional private agents can make relevant forecasting errors about the unobservables. Then, more transparent communication from the central bank's side can be helpful in narrowing the gap between the expected and actual values of the unobservables and ensure that the economy does not significantly deviate from its equilibrium path. The model of this paper achieves a similar conclusion with a slightly different information hierarchy (fully informed public agents) and integrates the findings of Han (2024) by paying attention to both the problems of equilibrium determinacy in the presence of alternative communication strategies and the phenomenon of the "Paradox of Transparency", which is neglected by Han (2024).

Since it benefits from greater behavioural realism (Evans & Honkapohja, 2009), the adaptive learning approach employed by Han (2024) may be seen as more appropriate than the traditional rational expectations modelling of this paper. However, as emphasized by Eusepi & Preston (2018) in their survey of the literature, learning can limit the set of policies available to central banks, which makes it possible that the New Keynesian models incorporating AL do not reproduce all the possible equilibrium outcomes. In fact, the complexity of the expectation formation process under AL can lead to either unstable equilibria or equilibria which are possible in principle but inconsistent with the central bank's target.

The result that, in a rational expectations, New Keynesian model with fully informed public agents and partially informed households and firms, the satisfaction of the Taylor principle is not a strictly necessary condition to establish equilibrium determinacy has been previously achieved by Sorge & Vota (2025), who are interested in studying the equilibrium characterization when reversing the information hierarchy posed by Lubik et al. (2023). The present manuscript proves that, at least for the full information case, this key theoretical prediction is robust to accounting for some elements that are absent in Sorge & Vota (2025), such as the persistence of exogenous shocks, fiscal policy, labour market dynamics, institutional communication with diverse degrees of transparency, and news shocks. On the other hand, the analysis reported in the previous Section 5 departs from the findings of Sorge & Vota (2025), according to which asymmetric information delivers multiple linear sunspot equilibria, by showing that instead it possibly involves equilibrium inexistence. This is an obvious consequence of explicitly accounting for institutional communication and news shocks.

As concerns the emergence of the "Paradox of Transparency", Sánchez (2013) finds that this phenomenon is consequential to the disclosure of the central bank's preferences about inflation stabilization and economic growth and the related strategic interaction between the central bank itself and the private sector. This paper makes a step forward by demonstrating that even when the preferences of the central bank (which are expressed by the policy coefficients of the Taylor rule) are perfectly known to households and firms as postulated by the rational expectations paradigm, the Paradox equally arises as a result of the noisy content of the informative signal.

In light of the theoretical analysis performed in this research work, it is possible to argue that the dominant individual-based approach is not suitable for evaluating the entity and determinants of job insecurity, as this latent variable is affected by macroeconomic shocks displaying some degree of persistence rather than purely random, unpredictable idiosyncratic disturbances.

Even the few attempts of evaluating the marginal effects of macroeconomic variables on job insecurity by non-structural, unrestricted empirical models performed until today (Ellonen & Nätti, 2015; Johnston et al., 2020) can lead to biased conclusions, as such settings ignore the persistence of exogenous shocks, the informative channel of fiscal and monetary policy, the long-run, optimizing behaviour of households and firms, and the possible equilibrium inexistence of the (realistic and common) asymmetric information setting. This entails that structural models can safely replicate the path of the main macroeconomic aggregates only in the rare and temporary cases of perfect communication transparency from policymakers, making their relevance for policy purposes questionable in the more realistic case of asymmetric information.

Finally, this manuscript provides the supporters of policy transparency with an additional argument in favour of their position. Some established contributions to the literature stress the importance of transparency in decreasing actual inflation and keeping inflation pressure under control (mainly thanks to the credibility gained by the monetary policy authority through a clear and exhaustive communication activity and the revelation of preferences for output and price stability), while others assert that opaqueness is preferable to achieve the price stability objective (Weber, 2016). In a similar vein, some research studies show that transparency about fiscal policy goals enhances economic efficiency and reduces uncertainty (Alt & Lassen, 2006; Arbatli & Escolano, 2015; Arapis & Reitano, 2018; De Simone et al., 2019), whereas others warn against the common wisdom by underscoring that fiscal transparency narrows the margins for policy flexibility because of the disincentive of the government to reconsider the promises made to the public (Heald, 2003).

The present paper documents that the economic consequences of implementing a not fully transparent communication strategy are possibly dramatic, even when the government is bound to fiscal balance and the central bank does not have a definite preference for output or inflation stability.

## 7. A possible extension to the mature workers' context

Although not explicitly incorporating anagraphic heterogeneity, the model developed in this manuscript can be used to investigate the determinants of the job insecurity of mature workers (another relevant research avenue) under the plausible assumption that, thanks to the stationarity property of the exogenous variables and shocks, the white noise nature of individual disturbances, and the uncertainty minimization role of the fully transparent institutional communication, the common knowledge equilibrium paths of diverse age groups are qualitatively similar. In other words, the job insecurity of mature workers is likely pinned down by the general equation (33), even if the dimension of the reduced-form coefficients probably reflects age-specific characteristics of this group (like lower intertemporal elasticity of substitution, lower marginal propensity to consume, and higher responsiveness to news shocks) that can substantially contribute to the expectations formation process about the deviation of the actual unemployment rate and other endogenous variables from their respective steady-state values.

The insight above can be empirically assessed by: (*a*) validating the New Keynesian model on both aggregate data and data related to mature workers after properly calibrating it, (*b*) estimating job insecurity both for the whole economy and mature workers as in equation (33) through the parameters estimated /and or selected for the calibration purposes, (*c*) performing the Johansen test for cointegration on the two estimated time series (aggregate job insecurity and mature workers job insecurity), and (*d*) testing for the null hypothesis that the slope and vertical intercept of a regression model whose dependent variable is the aggregate job insecurity and unique covariate is the job insecurity of mature workers are, respectively, equal to one and zero.

If the model displays a good fit of the mature workers' data too, the two estimated job insecurity time series (that for the whole economy and that of mature workers) are cointegrated, and the slope of the regression model is equal to one while the vertical intercept is null (which is a powerful indication of the lack of aggregate fallacy), then one can conclude that the reduced-form representation (20)-(33) is robust to anagraphic heterogeneity.

Upon successfully passing these checks, the job insecurity of mature workers estimated under the common knowledge assumption can be compared with the job insecurity measure extracted by the Survey of Health, Ageing, and Retirement in Europe (SHARE) conducted (with irregular frequency) between 2004 and 2022. This is a rich database that involves the responses of over 50 workers living in 28 European and non-European countries to questions about physical and mental health, healthcare, work and retirement, income and wealth, social networks, and socio-demographic and psychological aspects. The survey comprises, among others, a variable related to the risk subjectively perceived by over 50 workers about the interruption of their careers.

Despite the theoretical measure obtained by the theoretical model of this paper, the SHARE job insecurity proxy implicitly captures the equilibrium effect (if any) of non-fully-transparent institutional communication, as these data come from the real world where central banks and governments opt for an ordinary opaque communication strategy.

Evaluating vis-à-vis the theoretical job insecurity of mature workers under common knowledge and the actual job insecurity of mature workers taken from SHARE would allow researchers to set up an interesting counterfactual analysis about the impact of full institutional transparency on this relevant variable.

## 8. Concluding remarks

This manuscript made an attempt to contribute to the research field on job insecurity from a novel macroeconomic perspective as an alternative to the dominant individual-based approach.

More precisely, the paper tried to uncover the predictive capacity of key macroeconomic variables, with a particular focus on the role of information.

To accomplish this task, the paper proposed an original rational expectations New Keynesian model with asymmetric information in which partially informed private agents receive a noisy signal from their fully informed public counterparts.

The main finding of the research study was that equilibrium determinacy and uniqueness are achieved only in a common knowledge environment in which public agents disclose all the available information about the unobservables according to a fully transparent communication strategy. Along this equilibrium path, informative shocks can display either a positive or negative effect on the model's endogenous variables, potentially leading to the realization of the "Paradox of Transparency". Similarly, expectations (including job insecurity, defined as the expected deviation of the unemployment rate from its steady-state value) can either positively or negatively respond to the informative signal and shocks.

On the other hand, in the presence of opaque institutional communication, despite agents knowing the structure of the economy, equilibrium can potentially disappear, with dramatic consequences in terms of welfare for households and firms. This result indicates that, in the absence of relevant information on the state and perspectives of the economy, the entity of the macroeconomic uncertainty to which the agents are exposed is such that they can still form expectations about the unobservables, but they are eventually unable to efficiently allocate their resources in the long run.

In light of the above, policymakers should strive to provide households and firms with complete and clear informative feedback about the state of the economy, such that the private sector can make optimal decisions consistent with the achievement of the long-run equilibrium outcome. Otherwise, the market-clearing conditions can potentially be unsatisfied (notwithstanding the prices realized in the single markets), and both the single agents and the economy as a whole can face a potentially large loss in terms of key macroeconomic variables like consumption, saving, investment, unemployment, and national income. Thus, job insecurity can take two alternative forms: in the common knowledge setting, it is the subjective risk of being unemployed at a future time optimally estimated by private agents conditional to their available information set, whereas, in the asymmetric information case, it can become a component of the overall Knightian (i.e., unpredictable) economic uncertainty.

A natural extension of the current work can be assessing the sensitivity of the baseline results presented in this paper through the alternative adaptive learning modelling. This would allow researchers to discover the Expectational Stability (E-stability) equilibrium properties by following the research line on the expectationally driven business cycles traced by Dombeck (2022). In addition, estimating the coefficients of the reduced-form solution of the New Keynesian model under common knowledge enables researchers to empirically assess the entity of job insecurity in the theoretical case of fully transparent institutional communication. The comparison between this artificial construct and a proxy of actual job insecurity represents a valuable estimate of the loss in terms of welfare caused by opaque communication.

## Declarations

### Declaration of conflicting interest

Both authors have no conflicts of interest to disclose.

### Declarations Funding statement

This manuscript has received no financing.

### Ethical approval and consent statements

The authors' study involves neither human nor animal participants. Both authors consent to the article's publication.

**Data availability statement**
There are no data associated with this article.

**AI use**
The authors have used AI only for proofreading purposes.

**Acknowledgment**
This paper was produced within the framework of the project funded by Next Generation EU – "Age-It – Ageing well in an ageing society" (PE0000015), CUP B83C22004880006 - National Recovery and Resilience Plan (NRRP) – PE8 – Mission 4, Component 2, Intervention 1.3. The opinions expressed are solely those of the authors and do not necessarily reflect those of the European Union or the European Commission. Neither the European Union nor the European Commission can be held responsible for such contents.

# Appendix

**Table 1** Coefficients of the common knowledge solution for the actual interest rate

| Coefficient | Value |
|---|---|
| $z_0^r$ | $\dfrac{\sigma(s_0 c_1 - c_0 s_1)}{s_1 - \sigma[c_1(\gamma_2 + s_2) + \gamma_2 s_1]}$ |
| $z_1^r$ | $-\dfrac{\sigma\gamma_1\rho_{\bar{y}}^3(c_1 + s_1)}{(1 - \rho_{\bar{y}})\{s_1 - \sigma[c_1(\gamma_2 + s_2) + \gamma_2 s_1]\}}$ |
| $z_2^r$ | $-\dfrac{\sigma\gamma_1\rho_{\bar{y}}^2(c_1 + s_1)}{(1 - \rho_{\bar{y}})\{s_1 - \sigma[c_1(\gamma_2 + s_2) + \gamma_2 s_1]\}}$ |
| $z_3^r$ | $\dfrac{\sigma\rho_g[c_1(\gamma_3 - s_3) - c_3 s_1 + \gamma_3 s_1 - s_1]}{s_1 - \sigma[c_1(\gamma_2 + s_2) + \gamma_2 s_1]}$ |
| $z_4^r$ | $\dfrac{\sigma[c_1(\gamma_3 - s_3) - c_3 s_1 + \gamma_3 s_1 - s_1]}{s_1 - \sigma[c_1(\gamma_2 + s_2) + \gamma_2 s_1]}$ |
| $z_5^r$ | $\dfrac{\sigma\rho_{tax}[c_1(\gamma_4 - s_4) + s_1 c_4 + s_1\gamma_4]}{s_1 - \sigma[c_1(\gamma_2 + s_2) + \gamma_2 s_1]}$ |
| $z_6^r$ | $\dfrac{\sigma[c_1(\gamma_4 - s_4) + s_1 c_4 + s_1\gamma_4]}{s_1 - \sigma[c_1(\gamma_2 + s_2) + \gamma_2 s_1]}$ |
| $z_7^r$ | $-\dfrac{\sigma\rho_\chi(c_1 - s_1)(\gamma_5 - \varphi_2)}{s_1 - \sigma[c_1(\gamma_2 + s_2) + \gamma_2 s_1]}$ |
| $z_8^r$ | $-\dfrac{\sigma(c_1 - s_1)(\gamma_5 - \varphi_2)}{s_1 - \sigma[c_1(\gamma_2 + s_2) + \gamma_2 s_1]}$ |
| $z_9^r$ | $\dfrac{\sigma\varphi_1(c_1 - s_1)}{s_1 - \sigma[c_1(\gamma_2 + s_2) + \gamma_2 s_1]}$ |
| $z_{10}^r$ | $\dfrac{\sigma\varphi_3(c_1 - s_1)}{s_1 - \sigma[c_1(\gamma_2 + s_2) + \gamma_2 s_1]}$ |

**Table 2** Coefficients of the common knowledge solution for actual output

| Coefficient | Value |
|---|---|
| $z_0^y$ | $-\dfrac{(s_0 c_1 - c_0 s_1)}{s_1 - \sigma[c_1(\gamma_2 + s_2) + \gamma_2 s_1]}$ |
| $z_1^y$ | $\dfrac{\gamma_1\rho_{\bar{y}}^3(c_1 + s_1)}{(1 - \rho_{\bar{y}})\{s_1 - \sigma[c_1(\gamma_2 + s_2) + \gamma_2 s_1]\}}$ |
| $z_2^y$ | |

$$\frac{\gamma_1 \rho_{\bar{y}}^2 (c_1 + s_1)}{(1 - \rho_{\bar{y}})\{s_1 - \sigma[c_1(\gamma_2 + s_2) + \gamma_2 s_1]\}}$$

| | |
|---|---|
| $z_3^y$ | $-\dfrac{\rho_g c_1(\gamma_3 - s_3) - \rho_g c_3 s_1 + \rho_g \gamma_3 s_1 - \rho_g s_1}{s_1 - \sigma[c_1(\gamma_2 + s_2) + \gamma_2 s_1]}$ |
| $z_4^y$ | $-\dfrac{c_1(\gamma_3 - s_3) - c_3 s_1 + \gamma_3 s_1 - s_1}{s_1 - \sigma[c_1(\gamma_2 + s_2) + \gamma_2 s_1]}$ |
| $z_5^y$ | $-\dfrac{\rho_{tax} c_1(\gamma_4 - s_4) + \rho_{tax} s_1 c_4 + \rho_{tax} s_1 \gamma_4}{s_1 - \sigma[c_1(\gamma_2 + s_2) + \gamma_2 s_1]}$ |
| $z_6^y$ | $-\dfrac{c_1(\gamma_4 - s_4) + s_1 c_4 + s_1 \gamma_4}{s_1 - \sigma[c_1(\gamma_2 + s_2) + \gamma_2 s_1]}$ |
| $z_7^y$ | $\frac{[\sigma \rho_\chi(c_1 - s_1)(\gamma_5 - \varphi_2)][c_1(\gamma_2 + s_2) + \gamma_2 s_1] + \gamma_5 \rho_\chi(c_1 - s_1)\{s_1 - \sigma[c_1(\gamma_2 + s_2) + \gamma_2 s_1]\} - \varphi_2 \rho_\chi(c_1 - s_1)\{s_1 - \sigma[c_1(\gamma_2 + s_2) + \gamma_2 s_1]\}}{s_1^2 - s_1 \sigma[c_1(\gamma_2 + s_2) + \gamma_2 s_1]}$ |
| $z_8^y$ | $\frac{[\sigma(c_1 - s_1)(\gamma_5 - \varphi_2)][c_1(\gamma_2 + s_2) + \gamma_2 s_1] + \gamma_5(c_1 - s_1)\{s_1 - \sigma[c_1(\gamma_2 + s_2) + \gamma_2 s_1]\} - \varphi_2(c_1 - s_1)\{s_1 - \sigma[c_1(\gamma_2 + s_2) + \gamma_2 s_1]\}}{s_1^2 - s_1 \sigma[c_1(\gamma_2 + s_2) + \gamma_2 s_1]}$ |
| $z_9^y$ | $-\dfrac{\varphi_1(c_1 - s_1)}{s_1 - \sigma[c_1(\gamma_2 + s_2) + \gamma_2 s_1]}$ |
| $z_{10}^y$ | $-\dfrac{\varphi_3(c_1 - s_1)}{s_1 - \sigma[c_1(\gamma_2 + s_2) + \gamma_2 s_1]}$ |

**Table 3** Coefficients of the common knowledge solution for the output gap

| Coefficient | Value |
|---|---|
| $z_0^{\hat{y}}$ | $z_0^y$ |
| $z_1^{\hat{y}}$ | $(z_1^y - \rho_{\bar{y}}^2)$ |
| $z_2^{\hat{y}}$ | $(z_2^y - \rho_{\bar{y}})$ |
| $z_3^{\hat{y}}$ | $z_3^y$ |
| $z_4^{\hat{y}}$ | $z_4^y$ |
| $z_5^{\hat{y}}$ | $z_5^y$ |
| $z_6^{\hat{y}}$ | $z_6^y$ |
| $z_7^{\hat{y}}$ | $z_7^y$ |
| $z_8^{\hat{y}}$ | $z_8^y$ |
| $z_9^{\hat{y}}$ | $z_9^y$ |
| $z_{10}^{\hat{y}}$ | $z_{10}^y$ |

**Table 4** Coefficients of the common knowledge solution for the expected output gap

| Coefficient | Value |
|---|---|
| $z_0^{E\hat{y}}$ | $\rho_{\bar{y}} z_1^{\hat{y}}$ |
| $z_1^{E\hat{y}}$ | $\rho_{\bar{y}} z_1^{\hat{y}}$ |
| $z_2^{E\hat{y}}$ | $z_1^{\hat{y}}$ |
| $z_3^{E\hat{y}}$ | $\rho_g z_3^{\hat{y}}$ |
| $z_4^{E\hat{y}}$ | $z_3^{\hat{y}}$ |
| $z_5^{E\hat{y}}$ | $\rho_{tax} z_5^{\hat{y}}$ |
| $z_6^{E\hat{y}}$ | $z_5^{\hat{y}}$ |
| $z_7^{E\hat{y}}$ | $\rho_\chi z_7^{\hat{y}}$ |
| $z_8^{E\hat{y}}$ | $z_7^{\hat{y}}$ |

**Table 5** Coefficients of the common knowledge solution for the expected inflation rate

| Coefficient | Value |
|---|---|
| $z_0^{E\pi}$ | $\dfrac{z_0^y (\alpha_\pi k + \alpha_y + \sigma)}{1 - \alpha_\pi \beta}$ |
| $z_1^{E\pi}$ | $\dfrac{(z_1^y - \rho_{\bar{y}}^2)(\alpha_\pi k + \alpha_y + \sigma)}{1 - \alpha_\pi \beta}$ |
| $z_2^{E\pi}$ | $\dfrac{(z_2^y - \rho_{\bar{y}})(\alpha_\pi k + \alpha_y + \sigma)}{1 - \alpha_\pi \beta}$ |
| $z_3^{E\pi}$ | $\dfrac{z_3^y (\alpha_\pi k + \alpha_y + \sigma)}{1 - \alpha_\pi \beta}$ |
| $z_4^{E\pi}$ | $\dfrac{z_4^y (\alpha_\pi k + \alpha_y + \sigma)}{1 - \alpha_\pi \beta}$ |
| $z_5^{E\pi}$ | $\dfrac{z_5^y (\alpha_\pi k + \alpha_y + \sigma)}{1 - \alpha_\pi \beta}$ |
| $z_6^{E\pi}$ | $\dfrac{z_6^y (\alpha_\pi k + \alpha_y + \sigma)}{1 - \alpha_\pi \beta}$ |
| $z_7^{E\pi}$ | $\dfrac{z_7^y (\alpha_\pi k + \alpha_y + \sigma)}{1 - \alpha_\pi \beta}$ |
| $z_8^{E\pi}$ | $\dfrac{z_8^y(\alpha_\pi k + \alpha_y + \sigma)}{1 - \alpha_\pi \beta}$ |
| $z_9^{E\pi}$ | $\dfrac{z_9^y(\alpha_\pi k + \alpha_y + \sigma)}{1 - \alpha_\pi \beta}$ |
| $z_{10}^{E\pi}$ | $\dfrac{z_{10}^y(\alpha_\pi k + \alpha_y + \sigma)}{1 - \alpha_\pi \beta}$ |
| $z_{11}^{E\pi}$ | $-\dfrac{(\alpha_\pi k + \alpha_y)}{1 - \alpha_\pi \beta}$ |
| $z_{12}^{E\pi}$ | $\dfrac{\rho_\epsilon \alpha_\pi}{1 - \alpha_\pi \beta}$ |
| $z_{13}^{E\pi}$ | $\dfrac{\alpha_\pi}{1 - \alpha_\pi \beta}$ |

**Table 6** Coefficients of the common knowledge solution for the actual inflation rate

| Coefficient | Value |
|---|---|
| $z_0^\pi$ | $\beta z_0^{E\pi} + k z_0^{\bar{y}}$ |
| $z_1^\pi$ | $\beta z_1^{E\pi} + k z_1^{\bar{y}}$ |
| $z_2^\pi$ | $\beta z_2^{E\pi} + k z_2^{\bar{y}}$ |
| $z_3^\pi$ | $\beta z_3^{E\pi} + k z_3^{\bar{y}}$ |
| $z_4^\pi$ | $\beta z_4^{E\pi} + k z_5^{\bar{y}}$ |
| $z_5^\pi$ | $\beta z_5^{E\pi} + k z_5^{\bar{y}}$ |
| $z_6^\pi$ | $\beta z_6^{E\pi} + k z_6^{\bar{y}}$ |
| $z_7^\pi$ | $\beta z_7^{E\pi} + k z_7^{\bar{y}}$ |
| $z_8^\pi$ | $\beta z_8^{E\pi} + k z_8^{\bar{y}}$ |
| $z_9^\pi$ | $\beta z_9^{E\pi} + k z_9^{\bar{y}}$ |
| $z_{10}^\pi$ | $\beta z_{10}^{E\pi} + k z_{10}^{\bar{y}}$ |
| $z_{11}^\pi$ | $\beta z_{11}^{E\pi} - k$ |
| $z_{12}^\pi$ | $\beta z_{12}^{E\pi}$ |
| $z_{13}^\pi$ | $\beta z_{13}^{E\pi}$ |

**Table 7** Coefficients of the common knowledge solution for the household consumption

| Coefficient | Value |
|---|---|
| $z_0^c$ | $-\dfrac{s_0 s_1 c_1 - s_0 c_1^2 \sigma s_2 - s_0 \sigma \gamma_2 s_1 - c_0 s_1^2 + c_0 s_1 c_1 \sigma s_2 + c_0 \sigma \gamma_2 s_1^2}{s_1^2 - \sigma s_1[c_1(\gamma_2 + s_2) + \gamma_2 s_1]}$ |
| $z_1^c$ | $\dfrac{\sigma c_1 \gamma_1 \rho_{\bar{y}}^3 (c_1 + s_1)(\gamma_2 + s_2) + \gamma_1 c_1 \rho_{\bar{y}}^3 \{s_1 - \sigma[c_1(\gamma_2 + s_2) + \gamma_2 s_1]\}}{s_1(1 - \rho_{\bar{y}})\{s_1 - \sigma[c_1(\gamma_2 + s_2) + \gamma_2 s_1]\}}$ |
| $z_2^c$ | $\dfrac{\sigma c_1 \gamma_1 \rho_{\bar{y}}^2 (c_1 + s_1)(\gamma_2 + s_2) + \gamma_1 c_1 \rho_{\bar{y}}^2 \{s_1 - \sigma[c_1(\gamma_2 + s_2) + \gamma_2 s_1]\}}{s_1(1 - \rho_{\bar{y}})\{s_1 - \sigma[c_1(\gamma_2 + s_2) + \gamma_2 s_1]\}}$ |
| $z_3^c$ | $-\dfrac{\sigma c_1 \rho_g (\gamma_2 + s_2)[c_1(\gamma_3 - s_3) - c_3 s_1 + \gamma_3 s_1 - s_1] + \rho_g[c_1(\gamma_3 - s_3) - c_3 s_1]\{s_1 - \sigma[c_1(\gamma_2 + s_2) + \gamma_2 s_1]\}}{s_1^2 - \sigma s_1[c_1(\gamma_2 + s_2) + \gamma_2 s_1]}$ |
| $z_4^c$ | $-\dfrac{\sigma c_1 (\gamma_2 + s_2)[c_1(\gamma_3 - s_3) - c_3 s_1 + \gamma_3 s_1 - s_1] + \{s_1 - \sigma[c_1(\gamma_2 + s_2) + \gamma_2 s_1]\}[c_1(\gamma_3 - s_3) - c_3 s_1]}{s_1^2 - \sigma s_1[c_1(\gamma_2 + s_2) + \gamma_2 s_1]}$ |
| $z_5^c$ | $-\dfrac{\sigma c_1 \rho_{tax}(\gamma_2 + s_2)[c_1(\gamma_4 - s_4) + s_1 c_4 + s_1 \gamma_4] + \rho_{tax}[c_1(\gamma_4 - s_4) + s_1 c_4]\{s_1 - \sigma[c_1(\gamma_2 + s_2) + \gamma_2 s_1]\}}{s_1^2 - \sigma s_1[c_1(\gamma_2 + s_2) + \gamma_2 s_1]}$ |
| $z_6^c$ | $-\dfrac{\sigma c_1 (\gamma_2 + s_2)[c_1(\gamma_4 - s_4) + s_1 c_4 + s_1 \gamma_4] + [c_1(\gamma_4 - s_4) + s_1 c_4]\{s_1 - \sigma[c_1(\gamma_2 + s_2) + \gamma_2 s_1]\}}{s_1^2 - s_1 \sigma[c_1(\gamma_2 + s_2) + \gamma_2 s_1]}$ |
| $z_7^c$ | $\dfrac{\sigma \rho_\chi c_1 (\gamma_2 + s_2)(c_1 - s_1)(\gamma_5 - \varphi_2) - \rho_\chi[\varphi_2(c_1 - s_1) - c_1 \gamma_5]\{s_1 - \sigma[c_1(\gamma_2 + s_2) + \gamma_2 s_1]\}}{s_1^2 - \sigma s_1[c_1(\gamma_2 + s_2) + \gamma_2 s_1]}$ |
| $z_8^c$ | $\dfrac{\sigma c_1 (\gamma_2 + s_2)(c_1 - s_1)(\gamma_5 - \varphi_2) + [c_1 \gamma_5 - \varphi_2(c_1 - s_1)]\{s_1 - \sigma[c_1(\gamma_2 + s_2) + \gamma_2 s_1]\}}{s_1^2 - \sigma s_1[c_1(\gamma_2 + s_2) + \gamma_2 s_1]}$ |
| $z_9^c$ | $-\dfrac{\sigma \varphi_1 c_1 (\gamma_2 + s_2)(c_1 - s_1) + \varphi_1(c_1 - s_1)\{s_1 - \sigma[c_1(\gamma_2 + s_2) + \gamma_2 s_1]\}}{s_1^2 - \sigma s_1[c_1(\gamma_2 + s_2) + \gamma_2 s_1]}$ |
| $z_{10}^c$ | $-\dfrac{\sigma \varphi_3 c_1 (\gamma_2 + s_2)(c_1 - s_1) + \varphi_3(c_1 - s_1)\{s_1 - \sigma[c_1(\gamma_2 + s_2) + \gamma_2 s_1]\}}{s_1^2 - \sigma s_1[c_1(\gamma_2 + s_2) + \gamma_2 s_1]}$ |

**Table 8** Coefficients of the common knowledge solution for investment

| Coefficient | Value |
|---|---|
| $z_0^I$ | $-\dfrac{\sigma \gamma_2 (s_0 c_1 - c_0 s_1)}{s_1 - \sigma[c_1(\gamma_2 + s_2) + \gamma_2 s_1]}$ |
| $z_1^I$ | $\dfrac{\gamma_1 \rho_{\bar{y}}^3 \{s_1 - \sigma[c_1(\gamma_2 + s_2) + \gamma_2 s_1]\} + \sigma \gamma_1 \gamma_2 \rho_{\bar{y}}^3 (c_1 + s_1)}{(1 - \rho_{\bar{y}})\{s_1 - \sigma[c_1(\gamma_2 + s_2) + \gamma_2 s_1]\}}$ |
| $z_2^I$ | $\dfrac{\gamma_1 \rho_{\bar{y}}^2 \{s_1 - \sigma[c_1(\gamma_2 + s_2) + \gamma_2 s_1]\} + \sigma \gamma_1 \gamma_2 \rho_{\bar{y}}^2 (c_1 + s_1)}{(1 - \rho_{\bar{y}})\{s_1 - \sigma[c_1(\gamma_2 + s_2) + \gamma_2 s_1]\}}$ |
| $z_3^I$ | $-\dfrac{\gamma_2 \{\sigma \rho_g [c_1(\gamma_3 - s_3) - c_3 s_1 + \gamma_3 s_1 - s_1]\} + \gamma_3 \rho_g \{s_1 - \sigma[c_1(\gamma_2 + s_2) + \gamma_2 s_1]\}}{s_1 - \sigma[c_1(\gamma_2 + s_2) + \gamma_2 s_1]}$ |

$z_4^I$ $$-\frac{\sigma\gamma_2[c_1(\gamma_3 - s_3) - c_3 s_1 + \gamma_3 s_1 - s_1] + \gamma_3\{s_1 - \sigma[c_1(\gamma_2 + s_2) + \gamma_2 s_1]\}}{s_1 - \sigma[c_1(\gamma_2 + s_2) + \gamma_2 s_1]}$$

$z_5^I$ $$-\frac{\sigma\gamma_2\rho_{tax}[c_1(\gamma_4 - s_4) + s_1 c_4 + s_1\gamma_4] + \gamma_4\rho_{tax}\{s_1 - \sigma[c_1(\gamma_2 + s_2) + \gamma_2 s_1]\}}{s_1 - \sigma[c_1(\gamma_2 + s_2) + \gamma_2 s_1]}$$

$z_6^I$ $$-\frac{\sigma\gamma_2[c_1(\gamma_4 - s_4) + s_1 c_4 + s_1\gamma_4] + \gamma_4\{s_1 - \sigma[c_1(\gamma_2 + s_2) + \gamma_2 s_1]\}}{s_1 - \sigma[c_1(\gamma_2 + s_2) + \gamma_2 s_1]}$$

$z_7^I$ $$\frac{\sigma\gamma_2\rho_\chi(c_1 - s_1)(\gamma_5 - \varphi_2) + \gamma_5\rho_\chi\{s_1 - \sigma[c_1(\gamma_2 + s_2) + \gamma_2 s_1]\}}{s_1 - \sigma[c_1(\gamma_2 + s_2) + \gamma_2 s_1]}$$

$z_8^I$ $$\frac{\sigma\gamma_2(c_1 - s_1)(\gamma_5 - \varphi_2) + \gamma_5\{s_1 - \sigma[c_1(\gamma_2 + s_2) + \gamma_2 s_1]\}}{s_1 - \sigma[c_1(\gamma_2 + s_2) + \gamma_2 s_1]}$$

$z_9^I$ $$-\frac{\sigma\gamma_2\varphi_1(c_1 - s_1)}{s_1 - \sigma[c_1(\gamma_2 + s_2) + \gamma_2 s_1]}$$

$z_{10}^I$ $$-\frac{\sigma\gamma_2\varphi_3(c_1 - s_1)}{s_1 - \sigma[c_1(\gamma_2 + s_2) + \gamma_2 s_1]}$$

**Table 9** Coefficients of the common knowledge solution for the policy rate

| Coefficient | Value |
| --- | --- |
| $z_0^i$ | $\alpha_\pi z_0^\pi + \alpha_y z_0^{\hat{y}}$ |
| $z_1^i$ | $\alpha_\pi z_1^\pi + \alpha_y z_1^{\hat{y}}$ |
| $z_2^i$ | $\alpha_\pi z_2^\pi + \alpha_y z_2^{\hat{y}}$ |
| $z_3^i$ | $\alpha_\pi z_3^\pi + \alpha_y z_3^{\hat{y}}$ |
| $z_4^i$ | $\alpha_\pi z_4^\pi + \alpha_y z_4^{\hat{y}}$ |
| $z_5^i$ | $\alpha_\pi z_5^\pi + \alpha_y z_5^{\hat{y}}$ |
| $z_6^i$ | $\alpha_\pi z_6^\pi + \alpha_y z_6^{\hat{y}}$ |
| $z_7^i$ | $\alpha_\pi z_7^\pi + \alpha_y z_7^{\hat{y}}$ |
| $z_8^i$ | $\alpha_\pi z_8^\pi + \alpha_y z_8^{\hat{y}}$ |
| $z_9^i$ | $\alpha_\pi z_9^\pi + \alpha_y z_9^{\hat{y}}$ |
| $z_{10}^i$ | $\alpha_\pi z_{10}^\pi + \alpha_y z_{10}^{\hat{y}}$ |
| $z_{11}^i$ | $\alpha_\pi z_{11}^\pi + \alpha_y z_{11}^{\hat{y}}$ |
| $z_{12}^i$ | $\alpha_\pi z_{12}^\pi + \alpha_y z_{12}^{\hat{y}}$ |
| $z_{13}^i$ | $\alpha_\pi z_{13}^\pi + \alpha_y z_{13}^{\hat{y}}$ |

**Table 10** Coefficients of the common knowledge solution of the actual unemployment rate

| Coefficient | Value |
|---|---|
| $z_0^u$ | $-\theta z_0^y$ |
| $z_1^u$ | $-\theta\left(z_1^y - \rho_{\hat{y}}^2\right)$ |
| $z_2^u$ | $-\theta\left(z_2^y - \rho_{\hat{y}}\right)$ |
| $z_3^u$ | $-\theta z_3^y$ |
| $z_4^u$ | $-\theta z_4^y$ |
| $z_5^u$ | $-\theta z_5^y$ |
| $z_6^u$ | $-\theta z_6^y$ |
| $z_7^u$ | $-\theta z_7^y$ |
| $z_8^u$ | $-\theta z_8^y$ |
| $z_9^u$ | $-\theta z_9^y$ |
| $z_{10}^u$ | $-\theta z_{10}^y$ |
| $z_{11}^u$ | $\theta$ |
| $z_{12}^u$ | $\rho_u$ |

**Table 11** Coefficients of the common knowlegde solution for the job insecurity

| Coefficient | Value |
|---|---|
| $z_0^{Eu}$ | $z_0^u$ |
| $z_1^{Eu}$ | $\rho_{\hat{y}} z_1^u$ |
| $z_2^{Eu}$ | $z_1^u$ |
| $z_3^{Eu}$ | $\rho_g z_3^u$ |
| $z_4^{Eu}$ | $z_3^u$ |
| $z_5^{Eu}$ | $\rho_{tax} z_5^u$ |
| $z_6^{Eu}$ | $z_5^u$ |
| $z_7^{Eu}$ | $\rho_\chi z_7^u$ |
| $z_8^{Eu}$ | $z_7^u$ |
| $z_9^{Eu}$ | $\rho_u z_{12}^u$ |
| $z_{10}^{Eu}$ | $z_{12}^u$ |